\documentclass[twocolumn]{aastex631}
\usepackage{comment}
\usepackage{natbib} 
\usepackage{amsmath}
\usepackage[flushleft]{threeparttable}

\newcommand{\kms}{km s$^{-1}$}

\newcommand{\degree}{$^{\circ}$}

\defcitealias{Butterfield2024a}{FIREPLACE I}
\defcitealias{Butterfield2024b}{FIREPLACE II}
\defcitealias{Pare2024}{FIREPLACE III}

\begin{document}

\title[Filaments B-field]{\uppercase{ACES: The Magnetic Field in Large Filaments in the Galactic Center}}

\correspondingauthor{Dylan M. Par\'e}
\email{dylanpare@gmail.com}

\author[0000-0002-5811-0136]{Dylan M. Par\'e}
\affiliation{Joint ALMA Observatory, Alonso de Cordova 3107, Vitacura, Casilla 19001, Santiago de Chile, Chile}
\affiliation{National Radio Astronomy Observatory, 520 Edgemont Road, Charlottesville, VA 22903, USA}

\author[0009-0004-0121-1560]{Zi-Xuan Feng}
\affiliation{Università dell’Insubria, via Valleggio 11, 22100 Como, Italy}

\author[0000-0002-8455-0805]{Yue Hu}
\affiliation{Institute for Advanced Study, 1 Einstein Drive, Princeton, NJ 08540, USA}

\author[0000-0002-6362-8159]{Maya A. Petkova}
\affiliation{Space, Earth and Environment Department, Chalmers University of Technology, SE-412 96 Gothenburg, Sweden}

\author[0000-0002-2782-1082]{Jack Sullivan}
\affiliation{Department of Physics, University of Connecticut, 196A Auditorium Rd Unit 3046, Storrs, CT 06269}

\author[0000-0002-9483-7164]{Robin G. Tress}
\affiliation{Institute of Physics, Laboratory for Galaxy Evolution and Spectral Modelling, EPFL, Observatoire de Sauverny, Chemin Pegasi 51, 1290 Versoix, Switzerland}

\author[0000-0002-6073-9320]{Cara Battersby}
\affiliation{Department of Physics, University of Connecticut, 196A Auditorium Road, Unit 3046, Storrs, CT 06269, USA}

\author[0000-0001-5996-3600]{Janik Karoly}
\affiliation{Department of Physics and Astronomy, University College London, WC1E 6BT London, UK}

\author[0000-0002-7336-6674]{Alex Lazarian}
\affiliation{Department of Astronomy, University of Wisconsin-Madison, Madison, WI 53706, USA}

\author[0000-0002-5776-9473]{Dani Lipman}
\affiliation{Department of Physics, University of Connecticut, 196A Auditorium Rd Unit 3046, Storrs, CT 06269}

\author[0000-0003-1337-9059]{Xing Pan}
\affiliation{School of Astronomy and Space Science, Nanjing University, 163 Xianlin Avenue, Nanjing 210023, P.R.China}
\affiliation{Key Laboratory of Modern Astronomy and Astrophysics (Nanjing University), Ministry of Education, Nanjing 210023, P.R.China}
\affiliation{Center for Astrophysics $\vert$ Harvard \& Smithsonian, 60 Garden Street, Cambridge, MA, 02138, USA}

\author[0009-0008-2210-4931]{Marco Donati}
\affiliation{Università dell’Insubria, via Valleggio 11, 22100 Como, Italy}

\author[0000-0001-6113-6241]{Mattia C. Sormani}
\affiliation{Università dell’Insubria, via Valleggio 11, 22100 Como, Italy}

\author[0000-0001-8135-6612]{John Bally} 
\affiliation{Center for Astrophysics and Space Astronomy, Department of Astrophysical and Planetary Sciences, University of Colorado, Boulder, CO 80389, USA} 

\author[0000-0003-0410-4504]{Ashley~T.~Barnes}
\affiliation{European Southern Observatory (ESO), Karl-Schwarzschild-Stra{\ss}e 2, 85748 Garching, Germany}

\author[0000-0002-4013-6469]{Natalie O. Butterfield}
\affiliation{National Radio Astronomy Observatory, 520 Edgemont Road, Charlottesville, VA 22903, USA}

\author[0000-0001-8064-6394]{Laura Colzi}
\affiliation{Centro de Astrobiolog\'ia (CAB), CSIC-INTA, Carretera de Ajalvir km 4, Torrejón de Ardoz, 28850 Madrid, Spain}

\author[0000-0002-0706-2306]{Christoph Federrath}
\affiliation{Research School of Astronomy and Astrophysics, Australian National University, Canberra, ACT 2611, Australia}

\author[0000-0002-8586-6721]{Pablo Garcia}
\affiliation{Chinese Academy of Sciences South America Center for Astronomy, National Astronomical Observatories, CAS, Beijing 100101, China}
\affiliation{Instituto de Astronom\'ia, Universidad Cat\'olica del Norte, Av. Angamos 0610, Antofagasta, Chile}

\author[0000-0001-6431-9633]{Adam Ginsburg}
\affiliation{Department of Astronomy, University of Florida, 211 Bryant Space Science Center, P.O. Box 112055, Gainesville, FL 32611-2055, USA}

\author[0000-0002-1313-429X]{Savannah R. Gramze}
\affiliation{Department of Astronomy, University of Florida, Gainesville, FL 32611 USA}

\author[0000-0002-1730-8832]{Anika Schmiedeke}
\affiliation{Green Bank Observatory, PO Box 2, Green Bank, WV 24944, USA}

\author[0000-0002-7495-4005]{Christian Henkel}
\affiliation{Max-Planck-Institut f{\"u}r Radioastronomie, Auf dem H{\"u}gel 69, 53121 Bonn, Germany}

\author[0000-0001-9656-7682]{Jonathan D. Henshaw}
\affiliation{Astrophysics Research Institute, Liverpool John Moores University, IC2, Liverpool Science Park, 146 Brownlow Hill, Liverpool L3 5RF, UK}
\affiliation{Max Planck Institute for Astronomy, K\"{o}nigstuhl 17, D-69117 Heidelberg, Germany}

\author[0000-0002-3412-4306]{Paul T. Ho}
\affiliation{Institute of Astronomy and Astrophysics, Academia Sinica, 11F of ASMAB, AS/NTU No. 1, Sec. 4, Roosevelt Road, Taipei 10617, Taiwan}
\affiliation{East Asian Observatory, 660 N. A'ohoku, Hilo, Hawaii, HI 96720, USA}

\author[0000-0001-9155-3978]{Pei-Ying Hsieh}
\affiliation{National Astronomical Observatory of Japan, 2-21-1 Osawa, Mitaka, Tokyo 181-8588, Japan}

\author[0000-0003-4493-8714]{Izaskun Jimenez-Serra}
\affiliation{Centro de Astrobiolog\'{i}a (CAB), INTA-CSIC, Carretera de Ajalvir km 4, Torrejón de Ardoz, 28850 Madrid, Spain}

\author[0000-0002-0560-3172]{Ralf S.\ Klessen}
\affiliation{Universit\"{a}t Heidelberg, Zentrum f\"{u}r Astronomie, Institut f\"{u}r Theoretische Astrophysik, Albert-Ueberle-Str.\ 2, 69120 Heidelberg, Germany}
\affiliation{Universit\"{a}t Heidelberg, Interdisziplin\"{a}res Zentrum f\"{u}r Wissenschaftliches Rechnen, Im Neuenheimer Feld 225, 69120 Heidelberg, Germany}

\author[0000-0002-8804-0212]{J.~M.~Diederik Kruijssen}
\affiliation{Cosmic Origins Of Life (COOL) Research DAO, \href{https://coolresearch.io}{https://coolresearch.io}}

\author[0000-0001-6353-0170]{Steven N. Longmore}
\affiliation{Astrophysics Research Institute, Liverpool John Moores University, 146 Brownlow Hill, Liverpool L3 5RF, UK}
\affiliation{Cosmic Origins Of Life (COOL) Research DAO, \href{https://coolresearch.io}{https://coolresearch.io}}

\author[0000-0003-2619-9305]{Xing Lu}
\affiliation{Shanghai Astronomical Observatory, Chinese Academy of Sciences, 80 Nandan Road, Shanghai 200030, P.\ R.\ China}
\affiliation{State Key Laboratory of Radio Astronomy and Technology, A20 Datun Road, Chaoyang District, Beijing, 100101, P.\ R.\ China}

\author[0000-0001-8782-1992]{Elisabeth A.C. Mills}
\affiliation{Department of Physics and Astronomy, University of Kansas, 1251 Wescoe Hall Drive, Lawrence, KS 66045, USA}

\author[0000-0002-3078-9482]{\'Alvaro~S\'anchez-Monge}
\affiliation{Institut de Ci\`encies de l'Espai (ICE), CSIC, Campus UAB, Carrer de Can Magrans s/n, E-08193, Bellaterra, Barcelona, Spain}
\affiliation{Institut d'Estudis Espacials de Catalunya (IEEC), E-08860, Castelldefels, Barcelona, Spain}

\author[0000-0001-7330-8856]{Daniel L. Walker}
\affiliation{UK ALMA Regional Centre Node, Jodrell Bank Centre for Astrophysics, Oxford Road, The University of Manchester, Manchester M13 9PL, UK}

\author[0009-0002-7459-4174]{Jennifer Wallace}
\affiliation{Department of Physics, University of Connecticut, 196A Auditorium Road, Unit 3046, Storrs, CT 06269, USA}

\author[0000-0003-2384-6589]{Qizhou Zhang}
\affiliation{Center for Astrophysics $\vert$ Harvard \& Smithsonian, 60 Garden Street, Cambridge, MA, 02138, USA}

\begin{abstract}
   The Galactic Center (GC) is an extreme region of the Milky Way that is host to a complex set of thermal and non-thermal structures. In particular, the GC contains high-density gas and dust that is collectively referred to as the Central Molecular Zone (CMZ). In this work, we study a subset of HNCO filaments identified in band 3 ALMA observations of the GC obtained by the ALMA CMZ Exploration Survey (ACES) that are comparable to high density filaments identified in the Galactic Disk. We compare the orientation of the magnetic field derived from 214~$\mu$m SOFIA and 850~$\mu$m JCMT observations with the filament orientation to deftermine which mechanisms dominate the formation of these filaments. We observe a large range of magnetic orientations in our observed filaments indicating the complex environments the filaments are located in. We also compare the observational results to synthetic data sets created using an MHD model of the GC. Our analysis reveals that the dominant mechanisms local to the HNCO filaments vary throughout the GC with some filaments being dominated by supersonic turbulence and others by subsonic turbulence. The comparison to synthetic observations indicates that the observed filaments are in magnetically dominated environments that could be supporting these filaments against collapse. Our results on the CMZ filaments are also compared to results obtained on similar filaments located in the Galactic Disk, and we find that the filaments studied here are possible CMZ analogs to the dense filamentary ``bones'' observed previously in the Galactic Disk.
\end{abstract}


\section{INTRODUCTION} \label{sec:intro}
\begin{figure*}
    \centering
    \includegraphics[width=1.0\textwidth]{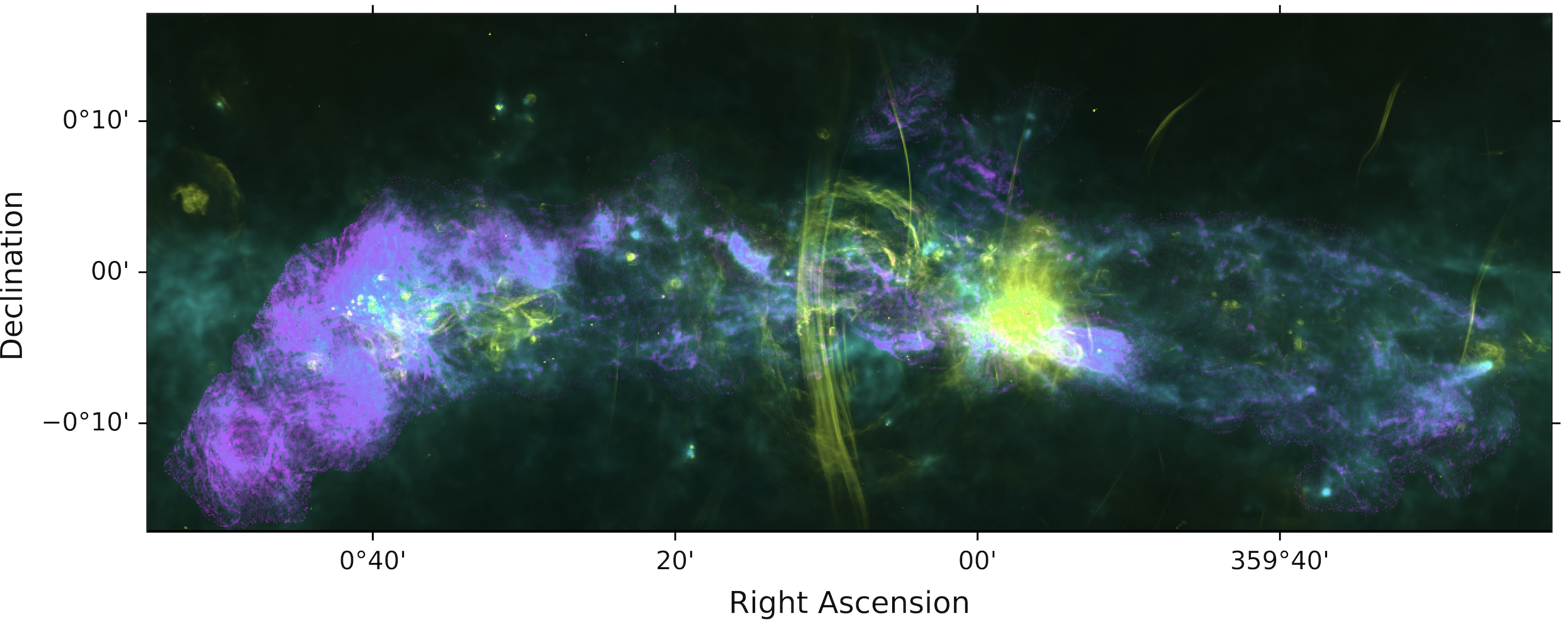}
    \caption{A 3-color view of the GC with 20 cm (1.28 GHz) MeerKAT radio emission tracing hot plasma in yellow \citep{Heywood2022}, 250~$\mu$m cool dust observed by Herschel in cyan \citep{Molinari2011}, and the 3.4 mm (87.9 GHz) ACES HNCO 4-3 moment 0 distribution observed in ALMA band 3 in purple (Longmore et al. submitted; Walker et al. submitted).}
    \label{fig:gc-3color}
\end{figure*}
The Galactic Center (GC) is the dynamic center of the Galaxy where material from the Galactic Disk is channeled inward along the Galactic Bar \citep{Sormani2019,Hatchfield2021,Tress2024}. The molecular material within radii of 100-300 pc in the GC is collectively referred to as the Central Molecular Zone (CMZ).

The CMZ is comprised of roughly 5\% of the total molecular gas in the Galaxy and 80\% of the dense molecular gas with column densities $>$10$^{23}$ cm$^{-2}$, with star formation rate estimates indicating it is 10\% that of the entire Milky Way \citep{Immer2012,Longmore2013,Barnes2017,Mills2017rev,Morris2023}. This star formation rate is lower than expected, however, given the high densities in the region \citep[see e.g.][for a recent review]{Henshaw2023}. The reason for the low CMZ SFR remains unclear, though there are multiple possible explanations such as the strength and driving of turbulence \citep{Kruijssen2014,Federrath2016a} or that the region is in a period of inactivity \citep{Kruijssen2014,Rathborne2014,Krumholz2015,Henshaw2023}. Alternatively, the strong magnetic field in the region could suppress star formation \citep[e.g.,][]{Morris1996b}. 

One way to enhance our understanding of what mechanism is regulating star formation in the CMZ is through the study of different morphological structures in the region. One such population of distinct morphological structures are molecular filaments. The largest of these filaments (with lengths $\geq$10 pc) in the Galactic plane are high density ($\sim$10$^{4}$ cm$^{-3}$) objects that trace the spiral structure of the Galaxy \citep{Reid2014,Goodman2014}. These structures are known as the ``bones,'' since they trace the highest density regions of the Milky Way spiral arms \citep{Zucker2015,Zucker2018}. The Stratospheric Observatory for Infrared Astronomy (SOFIA) legacy survey named: Filaments Extremely Long and Dark: A Magnetic Polarization Survey (FIELDMAPS) is analyzing the bones using polarimetric 214~$\mu$m High-resolution Airborne Wideband Camera Plus (HAWC+) observations to determine the role of the magnetic field in supporting the filaments against gravitational collapse \citep{Stephens2022,Coude2025}. FIELDMAPS finds that the magnetic field tends to be perpendicular to the bone G47.06+0.26 (G47) at its highest densities with a more complex field at lower densities \citep{Stephens2022}. More recently, FIELDMAPS presents results on a larger sample of 12 bones, finding that the field tends to be perpendicular in this larger sample as well \citep{Coude2025}. They conclude that the magnetic field is important in preventing cloud collapse against gravity for G47 and the larger bone sample in general. G47 and the other bones studied by FIELDMAPS are all well-removed from the GC \citep[Table 2 of][]{Coude2025}. It is therefore important to study whether filamentary structures with similar morphologies exist and, if so, if they exhibit similar magnetic field properties. The CMZ is the perfect region to search for similar filamentary structures because of its high densities and extreme properties.

Recently, HNCO filamentary structures similar to the bones were identified in Battersby et al. (submitted) using the ALMA CMZ Exploration Survey (ACES) observations. Battersby et al. (submitted) classified the HNCO filaments into Small-scale Filamentary structures (SFs) with lengths of $\sim$1 pc and Large-scale Filamentary structures (LFs) with lengths of $\sim$10 pc, similar lengths to the bones. Three of each type of filament were studied in detail in Battersby et al. (submitted), though several of each filament class are identified in the CMZ. The LFs observed in Battersby et al. (submitted) mainly trace large-scale orbital structures in the GC, similar to how the bones are found to trace the spiral arms of the Galaxy. The three LFs studied in Battersby et al. (submitted.) show a remarkable range of magnetic field orientation with respect to filament orientation, having variously a parallel, perpendicular, or mixed magnetic field alignment.

In this work we focus on an expanded subset of HNCO LFs to deepen our understanding of the connection between high density filaments and magnetic fields. Figure \ref{fig:gc-3color} shows a 3-color image of the complex distributions of thermal and non-thermal structures in the GC. The 1.28 GHz MeerKAT observations are shown in yellow which traces the hot plasma pervading the region \citep{Heywood2022}, the cyan shows the cool dust traced by the 250~$\mu$m Herschel-SPIRE observations \citep{Molinari2011} which reveals a figure-eight pattern that characterizes the large-scale structure of the CMZ \citep{Kruijssen2015,Walker2015,Battersby2025a,Battersby2025b,Walker2025,Lipman2025}, and the purple shows the ACES HNCO moment 0 emission which reveals the HNCO filaments (Longmore et al. submitted; Walker et al. submitted). The LFs studied in this work are extracted from the ACES HNCO distribution shown in Figure \ref{fig:gc-3color}, which largely traces the cool dust emission in the region at column densities above 10$^{22}$ cm$^{-2}$.

In Section \ref{sec:obs} we detail the observations used in this work. In Section \ref{sec:method} we describe how we isolate the HNCO LFs and how we study the alignment of the magnetic field with these structures. We then detail the properties of the MHD model we use to compare to our observations in Section \ref{sec:model} and the synthetic data that is created based on this model in Section \ref{sec:synth}. We present our results in Section \ref{sec:res}, discuss our findings in Section \ref{sec:disc} and summarize our results in Section \ref{sec:conc}.

\section{OBSERVATIONS} \label{sec:obs}
\begin{figure*}
    \centering
    \includegraphics[width=1.0\textwidth]{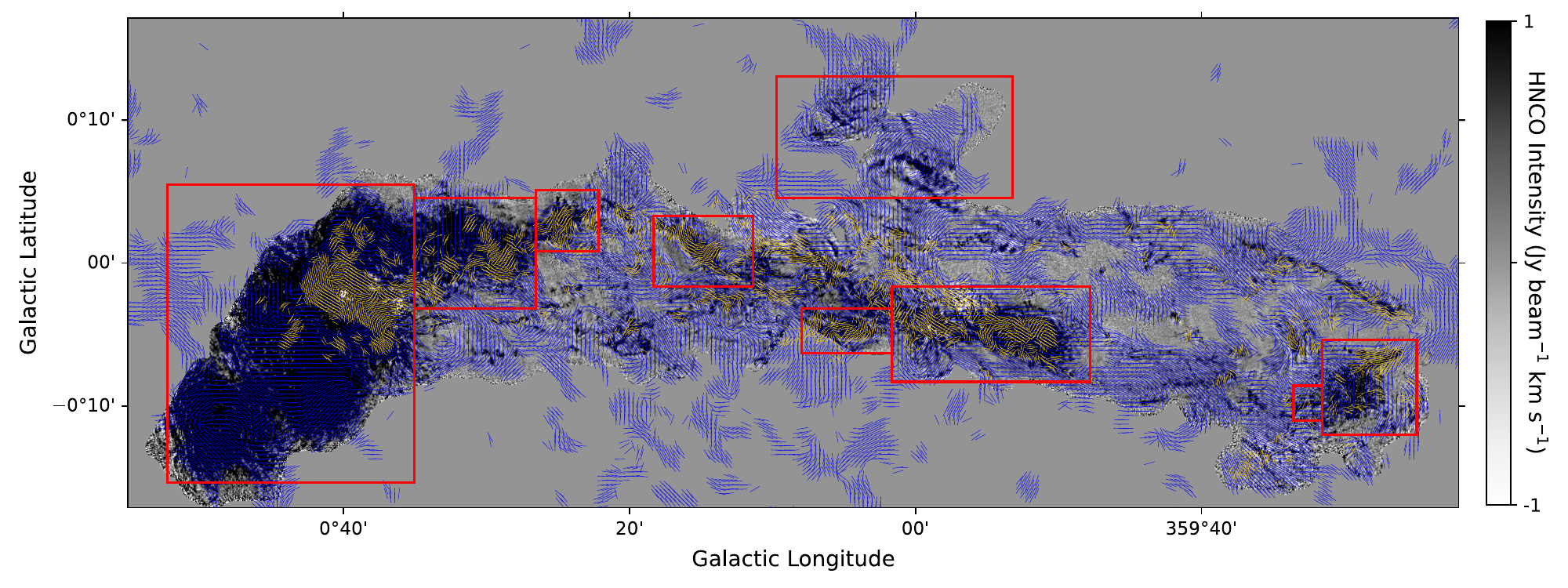}
    \caption{The ACES HNCO moment 0 distribution (Longmore et al. submitted) saturated to show the fainter filamentary features in the CMZ. FIREPLACE 214~$\mu$m and BISTRO 850~$\mu$m magnetic field orientations are overlaid as blue and yellow dashes, respectively \citep{Pare2024,karoly25}. The CMZ cloud regions and northern extent of the ACES coverage that have been masked out are indicated with red boxes.}
    \label{fig:hnco-bfield}
\end{figure*}
\subsection{ACES HNCO}
ACES is a cycle 8 band 3 large program on the Atacama Large Millimeter/sub-milimeter Array (ALMA, 2021.1.00172.S, PI: S. Longmore). This survey observed most of the molecular gas in the CMZ at 3mm with a 0.21 \kms\ velocity resolution (Longmore et al. submitted; Ginsburg et al. submitted; Walker et al. submitted; Lu et al. submitted; Hsieh et al. submitted) at an angular resolution of $\sim$3\arcsec \citep[or 0.1 pc at the assumed GC distance used throughout this work of 8.2 kpc,][]{Abuter2019}. We primarily use the ACES HNCO 4(0,4) -- 3(0,3) spectral line which has a frequency of 87925.237 MHz as described in detail in Walker et al. (submitted). This line is thought to trace dense molecular gas and low velocity shocks \citep{Martin2008,Henshaw2016a,Kelly2017}.

We display the moment 0 ACES distribution of the HNCO emission line in grayscale in Figure \ref{fig:hnco-bfield}, saturated to show the fainter HNCO filaments. We note that the negative emission in the HNCO map is a result of imaging artifacts caused by the edge of the ALMA primary beam and the imaging reconstruction method. The red boxes indicate the locations of prominent CMZ molecular clouds which are masked out to identify only filaments that are separate from molecular clouds as detailed in Section \ref{sec:method}. We also mask out the northern spur of the data set where structures may not be local to the CMZ, as shown with the northern-most red box. We note that there is a significant asymmetry in the ACES HNCO emission of the GC, where more HNCO emission is observed in the Eastern half of the GC. This asymmetry reflects the general asymmetry observed in the GC where a greater number of high density clouds are observed in the Eastern half of the GC (such as Sgr B2, the dust ridge, and the Brick) than in the Western half \citep[e.g.][]{Henshaw2023}.

\subsection{FIREPLACE Dust Polarization}
The Far-InfraREd Polarimetric Large Area CMZ Exploration (FIREPLACE) survey observed the cool dust in the CMZ at 214~$\mu$m using SOFIA/HAWC+. These observations have a pixel size of 4.9\arcsec\ \citep[beam size of 19.6\arcsec,][]{Butterfield2024a,Pare2024}. Since non-spherical grains preferentially align with their long axes perpendicular to the magnetic field orientation \citep[B-RAT alignment,][]{Lazarian07,Anderson2015}, the emission from FIREPLACE is polarized. This polarization enables the orientation of the magnetic field local to the grains to be inferred (by rotating the polarization angle by 90\degree).

We employ the same significance cuts used by \citet{Pare2024} to only consider the polarization on lines-of-sight that satisfy $I_{214}/\sigma_{I_{214}} >$ 200, $p_{214}/\sigma_{p_{214}} >$ 3.0, and $p_\%(214) <$ 50 where $I_{214}$ and $\sigma_{I_{214}}$ are the 214~$\mu$m emission and uncertainty, $p_{214}$ and $\sigma_{p_{214}}$ are the 214~$\mu$m polarization intensity and uncertainty, and $p_\%(214)$ is the percentage polarization ($p_{214}/I_{214}\times100)$. These significance cuts align with standard SOFIA/HAWC+ practices \citep{Harper2018}.

The plane of sky magnetic field orientation derived from the 214~$\mu$m FIREPLACE observations is shown as blue dashes in Figure \ref{fig:hnco-bfield}. The dashes shown are beam-sampled to the 19.6\arcsec\ FIREPLACE beam size and are shown with a constant length. 

The FIREPLACE magnetic field is argued to be mostly tracing CMZ structures since the magnetic field orientation largely traces the orientations of the CMZ clouds \citep{Pare2024,Pare2025}; however, we note that in the lower density regions of the CMZ the FIREPLACE magnetic field is possibly tracing a foreground structure in the central kpc of the Galaxy \citep{Pare2024}.

\subsection{BISTRO Dust Polarization}
The B-fields In STar-forming Region Observations (BISTRO) Survey \citep{dwt2017} observed the CMZ at 850\,$\mu$m \citep{karoly25} using the SCUBA-2/POL-2 instrument on the James Clerk Maxwell Telescope (JCMT). The JCMT has a beam size of 14.6\arcsec at 850\,$\mu$m \citep{mairs21} which corresponds to $\approx$0.6\,pc at 8.2\,kpc.

The details of the observations and data reduction are given in \citet{karoly25}. The data are reduced onto a 4\arcsec pixel grid and separate Stokes \textit{I}, \textit{Q} and \textit{U} maps are produced with an RMS of $\sim$ 10, 7, and 7 mJy\,beam$^{-1}$, respectively. These maps are then binned to 12\arcsec\ to compensate for the JCMT beam size and to match the magnetic field orientations presented in the BISTRO papers to facilitate comparison with those publications \citep[e.g.,][]{karoly25}.

We follow the selection criteria of \citet{karoly25} and analyze only the 12$\arcsec$ BISTRO vectors from lines of sight that satisfy: $I_{850}/\sigma_{I_{850}} >$ 50, $p_{850}/\sigma_{p_{850}} >$ 3.0, and $p_\%(850) <$ 20 where $I_{850}$ and $\sigma_{I_{850}}$ are the 850\,$\mu$m Stokes \textit{I} emission and uncertainty, $p_{850}$ and $\sigma_{p_{850}}$ are the 850~$\mu$m polarization intensity and uncertainty, and $p_\%(850)$ is the percentage polarization ($p_{850}/I_{850}\times100)$. The plane of sky magnetic field orientation derived from the 850~$\mu$m BISTRO observations at 12\arcsec\ resolution is shown as yellow dashes in Figure \ref{fig:hnco-bfield} with a constant length.

As with the FIREPLACE observations, the BISTRO observations are observed to largely trace the orientations of the CMZ clouds \citep{karoly25}. Furthermore, BISTRO lacks sensitivity to the fainter CMZ regions where the FIREPLACE magnetic field possibly traces a foreground magnetic field system. The BISTRO orientations studied in this work are therefore likely local to the CMZ rather than being some other field component along the line of sight. 

After implementing the FIREPLACE and BISTRO significance cuts described above we then need to associate polarization orientations to the specific LFs we study in this work. To do so we associate each significant orientation that is within one FIREPLACE or BISTRO beam size to an LF as an orientation associated with that LF. These sets of vectors are then used to determine the relative orientations of the polarization angles to the different LF structures, as presented in Section \ref{sec:res}. We note that because of the fewer JCMT vectors that are associated with the fainter HNCO structures, the JCMT observations are more restrictive and significant orientations are only obtained for the central three filament regions as indicated in Figure \ref{fig:fil_map}.

\section{METHODS} \label{sec:method}

\subsection{Identification of HNCO Filaments} \label{sec:meth-id}

\begin{figure*}
    \centering
    \includegraphics[width=1.0\linewidth]{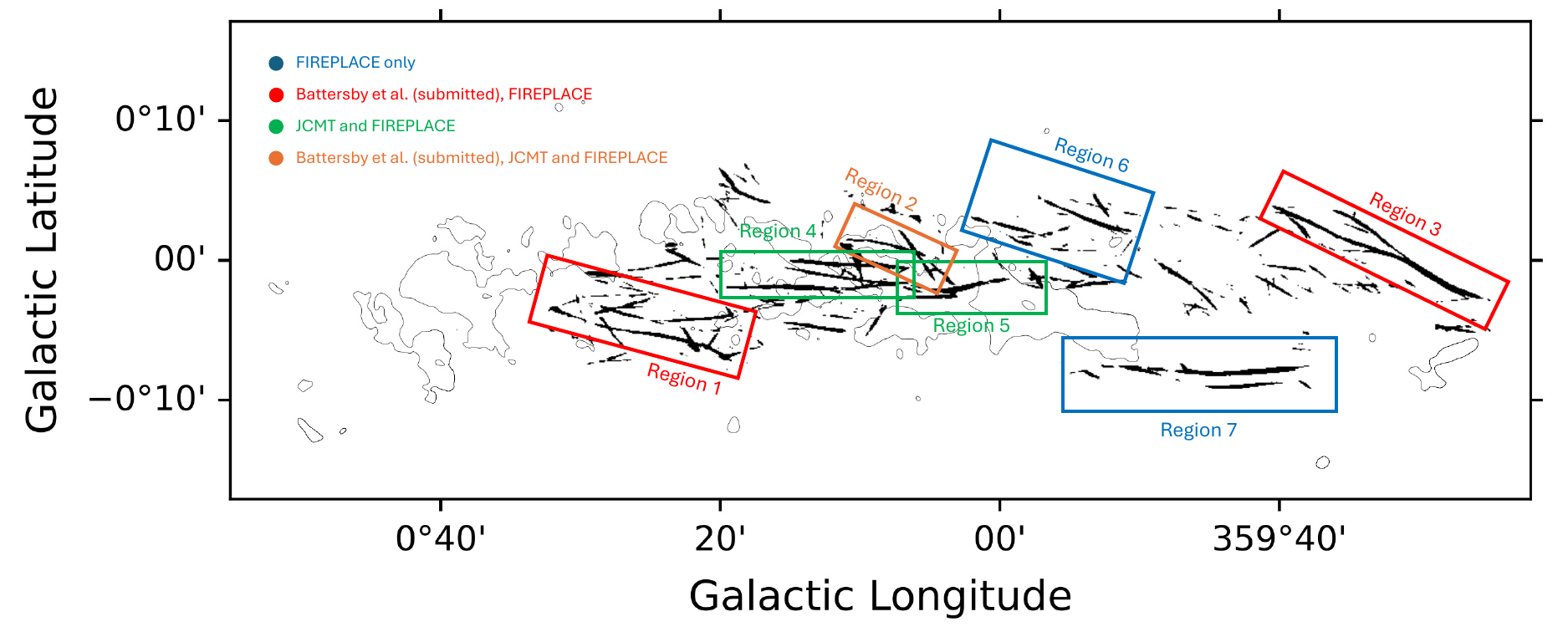}
    \caption{The final RHT distribution used to identify filamentary structures in the ACES HNCO distribution. We show the RHT distribution obtained from the HNCO distribution that has been convolved to the 19.6\arcsec\ FIREPLACE beam size. The black contour traces the cool dust traced by the FIREPLACE 214~$\mu$m observations at a level of 6 Jy pixel$^{-1}$ and indicates the CMZ molecular clouds that have been masked out. The boxes indicate the locations of the filaments that are targeted for a detailed analysis in this work. Blue boxes indicate filaments that coincide with  only FIREPLACE magnetic field orientations and green boxes indicate filaments that coincide with both FIREPLACE and BISTRO magnetic field orientations. In addition, red and orange boxes indicate the filaments previously studied in Battersby et al. (submitted) where the red boxes coincide with only FIREPLACE magnetic field orientations and the orange box coincides with both FIREPLACE and BISTRO magnetic field orientations. The filaments studied within each of these regions are shown in Figures \ref{fig:hro_rht_sofia_fig1} and \ref{fig:hro_rht_sofia_fig2}.}
    \label{fig:fil_map}
\end{figure*}

\begin{deluxetable*}{lccccccccc}
\tablecaption{HNCO LF Properties
\label{tab:hnco_prop}}
\tablewidth{0pt}
\tablehead{
\colhead{ID} & \colhead{Gal. Lon.} & \colhead{Gal. Lat.} &  \colhead{Length} & \colhead{Width} & \colhead{AR} & \colhead{HNCO Mom. 0} & \colhead{Vel.} & Vel. Disp. & \colhead{log$_{10}$ N(H$_2$)}
\\
  & (degree) & (degree) & (pc) & (pc) & & (Jy beam$^{-1}$ km s$^{-1}$) & (km s$^{-1}$) & (km s$^{-1}$) & (cm$^{-2}$)
}
\startdata
Region 1 & & & & & & & \\
LF 1a & 0.37 & -0.07 & 10.1 & 0.3 & 33.7 & 0.35 & 84 & 20 & 22.6 \\
LF 1b & 0.43 & -0.08 & 22.8 & 0.4 & 57.0 & 0.18 & 88 & 10 & 22.7 \\
\hline
Region 2 & & & & & & & \\
LF 2a & 0.10 & 0.00 & 7.7 & 0.4 & 19.3 & 1.80 & -10 & 25 & 22.5 \\
LF 2b & 0.13 & 0.02 & 9.8 & 0.5 & 19.6 & 1.20 & 75 & 28 & 22.6 \\
\hline
Region 3 & & & & & & & \\
LF 3 & 359.54 & 0.018 & 27.6 & 0.5 & 55.2 & 0.75 & -100 & 15 & 22.7 \\
\hline
Region 4 & & & & & & & \\
LF 4a & 0.18 & 0.00 & 14.4 & 0.6 & 24.0 & 0.76 & 28 & 40 & 22.6 \\
LF 4b & 0.22 & -0.03 & 18.4 & 0.4 & 46.0 & 0.57 & 25 & 35 & 22.7 \\
\hline
Region 5 & & & & & & & \\
LF 5 & 0.03 & -0.03 & 8.8 & 0.7 & 12.6 & 1.05 & -17 & 32 & 22.6 \\
\hline
Region 6 & & & & & & & \\
LF 6a & 359.90 & 0.05 & 12.0 & 0.5 & 24.0 & 0.30 & -60 & 20 & 22.4 \\
LF 6b & 359.98 & 0.03 & 9.4 & 0.5 & 18.8 & 0.52 & -30 & 30 & 22.4 \\
\hline
Region 7 & & & & & & & \\
LF 7a & 359.70 & -0.13 & 16.2 & 0.5 & 32.4 & 0.65 & -20 & 18 & 22.6 \\
LF 7b & 359.70 & -0.15 & 9.3 & 0.5 & 18.6 & 0.40 & -5 & 20 & 22.7 \\
\hline
\enddata
\tablecomments{For each row, the first column indicates the region (with regions shown in Figure \ref{fig:fil_map}) and the LFs within that region (as indicated in Figures \ref{fig:hro_rht_sofia_fig1} and \ref{fig:hro_rht_sofia_fig2}). The central Galactic longitude and latitude for each LF are listed in the second and third columns. The lengths and widths of each LF (in pc) are in the fourth and fifth columns, with the corresponding aspect ratio presented in the sixth column. A representative HNCO moment 0 intensity for each filament is displayed in the seventh column, with the average velocities and velocity dispersions listed in columns eight and nine. Column ten displays the average N(H$_2$) column density for each LF as derived from the Herschel dust continuum \citep{Molinari2011}.}
\end{deluxetable*}
To analyze the properties of the filaments with similar morphologies to those observed in the Galactic Disk and in Battersby et al. (submitted) we first masked out the HNCO emission that coincides with prominent CMZ molecular clouds. We identified the footprint of these prominent clouds using the SMA CMZoom 1.3 mm dust continuum observations of the GC \citep{Battersby2020} and using an $N(H_2)$ column density cut of $>1.0\times10^{23}$ cm$^{-2}$ to identify the densest clouds in the CMZ, where we used the column density derived from Herschel dust continuum observations of the GC \citep{Molinari2011,Mills2017}. This masking isolates the HNCO emission coinciding with the stream structures in the CMZ but that is distinct from the molecular clouds, which allows us to prioritize the longer LFs that are more comparable to the bones in the Galactic plane. This masked distribution is then convolved to the 12\arcsec\ and 19.6\arcsec\ beam sizes of the BISTRO and FIREPLACE data sets. We then apply the Rolling Hough Transform (RHT) algorithm to these convolved HNCO distributions to identify filamentary features \citep{Clark2014}. This identification builds on the work of Battersby et al. (submitted), by revealing more molecular filaments throughout the CMZ. The RHT parameters were chosen based on characteristic filament lengths and widths as determined in Battersby et al. (submitted). The window size was 540\arcsec\ (22 pc) and the smoothing diameter was 13\arcsec\ (0.5 pc). The RHT threshold parameter was set to 0.7. An RHT threshold parameter $<$1.0 allows the RHT to identify filaments that are physically coherent even if not connected visually. A range of RHT parameters was explored by varying the window size from 250\arcsec\ -- 750\arcsec\ ($\sim$10 -- 30 pc) and the smoothing diameter from 6\arcsec\ -- 51\arcsec\ ($\sim$0.25 -- 2.0 pc) and the above set of parameters was best able to extract the HNCO filamentary structures. 

Battersby et al. (submitted) noted the LFs are largely oriented parallel to the larger CMZ structure that is thought to be orbital streams or an ellipse that the molecular clouds travel along \citep{Kruijssen2015,Walker2025}. Battersby et al. (submitted) determined that the LFs are local to the GC using scale length relations and velocity profiles. The LFs are the particular filamentary structures we want to focus on in this work and they do not coincide with CMZ molecular clouds.

We identify the same filamentary features in the 12\arcsec\ BISTRO and 19.6\arcsec\ FIREPLACE resolutions. We keep both data sets at these separate resolutions to enable a beam-sampled comparison to both the FIREPLACE and BISTRO magnetic field orientations (presented in Section \ref{sec:res}). These filamentary structures are observed in the integrated velocity maps, ruling out the possibility of velocity crowding in thin velocity channel maps \citep{2023MNRAS.524.2994H}. We inspected the resulting filamentary distributions and removed any filaments at the edge of the ACES Field of View where the primary beam response of the ALMA observations falls below a normalized response of 0.2. The edge region of the ACES mosaic is dominated by noise, and filamentary features found there likely result from the elevated noise level at the edge of the ACES mosaic. The final RHT distribution obtained from this procedure is shown in Figure \ref{fig:fil_map}.  This figure also marks the extents of the masked molecular clouds as a black contour to indicate where we have masked the HNCO emission coinciding with the CMZ molecular clouds. Because of the masking of the clouds, there is an asymmetry in the distribution of LFs identified in the CMZ where more LFs are identified in the Western extent of the CMZ.

Numerous filamentary structures ($\sim$30) are identified from this procedure. In this work we choose to focus on the larger filaments (LFs) that have lengths $\gtrsim$10 pc in the distribution shown in Figure \ref{fig:fil_map}. These LFs are found within the regions marked with boxes in the figure, resulting in 12 LFs that are studied in detail in this work. The blue boxes indicate LFs coinciding with only FIREPLACE magnetic field orientations, while the green boxes indicate LFs coinciding with both FIREPLACE and BISTRO magnetic field orientations. The orange and red boxes indicate the LFs that were previously presented in Battersby et al. (2025, submitted). We inspected the 12 LFs in the HNCO velocity cubes to verify that they are coherent velocity structures and to determine the central velocities and velocity dispersions of each of these LFs. Table \ref{tab:hnco_prop} lists key properties of the LFs such as their lengths, widths, and velocities. We note that these properties are determined using the HNCO distributions that have been smoothed using the 13\arcsec\ RHT smoothing kernel. Analysis and discussion of these properties is presented in Section \ref{subsec:filaproper}, but we emphasize here that the LFs generally have lengths $\gtrsim$10 pc with velocity dispersions of $\sim$10 km s$^{-1}$ indicating that they are located in the GC. We note that three LFs studied in Battersby et al. (submitted) are part of our sample of LFs, where we match the numbering of the LFs in Battersby et al. (submitted), those being Regions 1, 2, and 3. We then number the remaining LFs from Galactic East to West.

\subsection{Alignment Measure} \label{sec:am}

To quantitatively analyze the relative orientations between the magnetic field and the LFs we apply the Histogram of Relative Orientation (HRO) method developed by \citet{Soler2013}. Originally designed to characterize the relative orientations between magnetic fields and density structures in MHD simulations \citep[e.g.,][]{Soler2013}, the HRO method has revealed that the relative orientation transits from parallel to perpendicular above a critical column density of 21.7 log$_{10}$ cm$^{-3}$ in Galactic Disk molecular structures \citep[e.g.,][]{PlanckXXXII2016, Soler2019, Fissel2019}. This transition reflects the changing balance between magnetic, turbulent kinetic, and gravitational energies \citep{Chen2016, Soler2017}. However, this transition is not observed in the GC, possibly indicating the dominance of shear in the CMZ which could prevent cloud collapse \citep{Federrath2016a,Pare2025}. In this study, we adapt the HRO method to compare the magnetic field orientation with that of the LFs, which can shed light on the role of magnetic fields in filament formation and evolution.

Following \citet{Clark2014}, the orientation of LFs identified by RHT can be determined by:
\begin{equation}
\langle\theta_\mathrm{RHT}\rangle = \frac{1}{2} \arctan\left[\frac{\int \sin(2\theta)R(\theta)d\theta}{\int \cos(2\theta)R(\theta)d\theta}\right],
\label{eq:rht_pa}
\end{equation}
where $R(\theta)$ is the RHT intensity as a function of angle $\theta$, calculated at each point along the LF. Then, the relative angle ($\phi$) between the LF and magnetic field is defined as:

\begin{equation}
    \phi=\sin^{-1}\left(\sin(|\langle\theta_\mathrm{RHT}\rangle-\theta_B|)\right), \label{eq:phi}
\end{equation}
where $\theta_B$ is the position angle of the magnetic field and $\phi\in[0, \pi/2]$. The use of $\sin^{-1}\sin$ in Equation \ref{eq:phi} ensures that $\phi$ is limited to the angle range 0 - 90\degree. We use a normalized parameter AM (alignment measure) introduced by \cite{Lazarian2018} to characterize the relative orientation (results presented in Section \ref{sec:res} and Table \ref{tab:bfield_prop}):
\begin{equation}
    \mathrm{AM}=\left\langle\cos 2\phi \right\rangle.
    \label{eq:am}
\end{equation}

The uncertainty in AM is calculated following \citet[][see their Appendix B]{Liu2023}:
\begin{multline}
\delta\mathrm{AM} = \sqrt{\left(\left\langle(\cos2\phi)^2\right\rangle - \mathrm{AM}^2 + \sum (2\sin(2\phi)\delta\phi)^2\right)/n^\prime},
\end{multline}
where $n^\prime$ is the number of independent data points within each intensity bin. Here, $\mathrm{AM}>0$ indicates that the magnetic field is parallel to the LF, $\mathrm{AM}<0$ indicates that the magnetic field is perpendicular to the LF, and $\mathrm{AM}\sim 0$ suggests no preferred orientation between the magnetic field and LF.

\section{MHD Model} \label{sec:model}
\begin{figure*}
    \centering
    \includegraphics[width=1.0\textwidth]{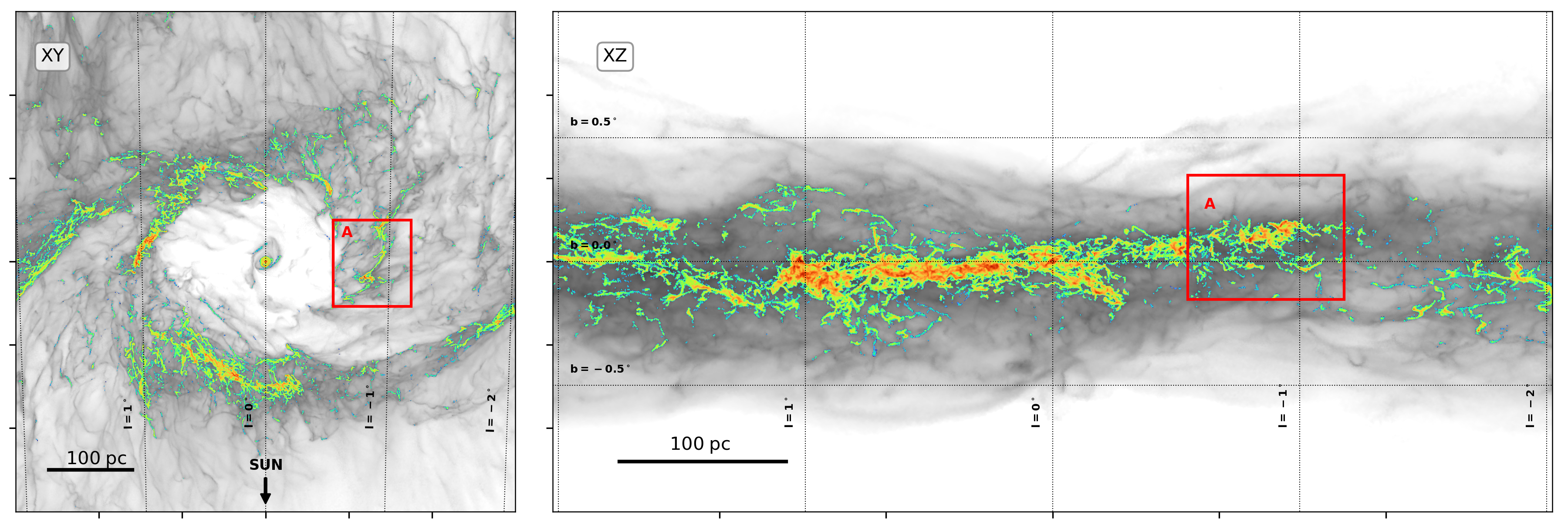}
    \caption{Face-on (left panel) and edge-on (right panel) H$_2$ column densities of the simulated CMZ from this work (Section \ref{sec:model}. The projections of the total H$_2$ gas is shown in gray-scale, while the dense ($n>10^3$~cm$^{-3}$) molecular gas is highlighted in color-scale. A zoom-in of box A can be seen in Figure \ref{fig:filament1}.}
    \label{fig:Simulations1}
\end{figure*}

\begin{figure}
    \centering
    \includegraphics[width=1.0\columnwidth]{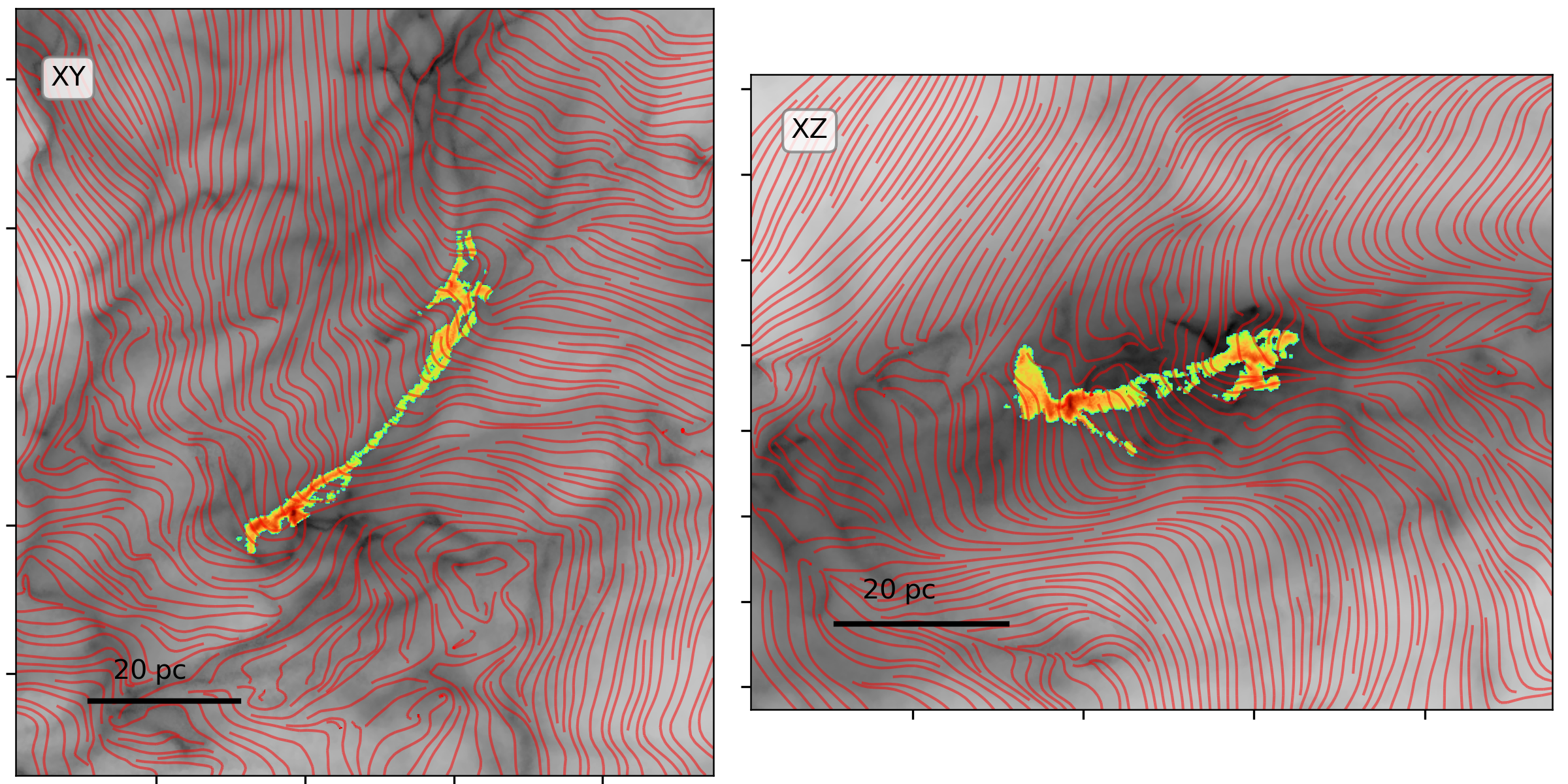}
    \caption{Example filament taken from Figure \ref{fig:Simulations1} as marked with the red box in that figure. Magnetic field lines are shown in red and are estimated by performing a mass-weighted average along the line of integration using the total gas density}. Gas clouds are stretched and sheared as they approach the apo-center of their orbit, generating their filamentary aspect.
    \label{fig:filament1}
\end{figure}

We performed a set of numerical MHD simulations to understand the dynamical origin of filamentary structure in the dense ISM of the CMZ. In particular, we are interested in exploring if and how the Galactic shear and the strong magnetic fields in the region can generate the LFs seen in observations. To this end, we designed simulations similar to the ones described in \citet{Tress2024}. 

We follow the ISM evolution in the entire barred region of the Milky Way, under the influence of an externally imposed barred background potential. Magnetic fields are included but the self-gravity of the gas is not considered. This way, the dynamical impact of the galactic potential and magnetic fields can be separated from star formation and feedback effects on the ISM, which instead can be addressed in a follow-up study. 

Compared to \citet{Tress2024}, the background potential and initial gas density profile are updated to represent better the Milky Way mass distribution. For the Galactic potential we take the model introduced by \citet{Hunter2024}, where the initial gas distribution of our MHD model follows their Equation 14. This results in a CMZ with a diameter of $\sim 300$~pc and a total mass of $\sim 2 \times 10^7$~M$_\odot$, in much better agreement with observations. 

To resolve the critical density of HNCO, a mass resolution of 1~M$_\odot$ is imposed for cells in the central 500~pc of the simulation domain (Lipman et al. submitted). To also study the inflow from the dust lanes, the mass resolution for cells in the barred region is set to 100~M$_\odot$. The rest of the domain is followed at a resolution of 1000~M$_\odot$. The central supermassive black hole Sgr A$^*$ is modeled as an accreting sink particle of mass $M_{\rm SMBH} = 4\times10^6$~M$_\odot$ and an accretion radius of $0.2$~pc. 

A gas density projection map of a snapshot of the simulation can be seen in Figure \ref{fig:Simulations1}. A plethora of filamentary and turbulent structures is observed in the dense molecular gas, highlighted in color-scale in the figure. A zoom-in onto one such structure is shown in Figure \ref{fig:filament1}, where the magnetic field has been estimated using a mass-weighted average from the total gas density. This example filament is formed as a molecular cloud which is sheared apart by the strong Galactic shear as it approaches apocenter of its x2 orbit around the Galactic center. 

\section{SYNTHETIC DATA} \label{sec:synth}
\begin{figure*}
    \centering
    \includegraphics[width=\linewidth]{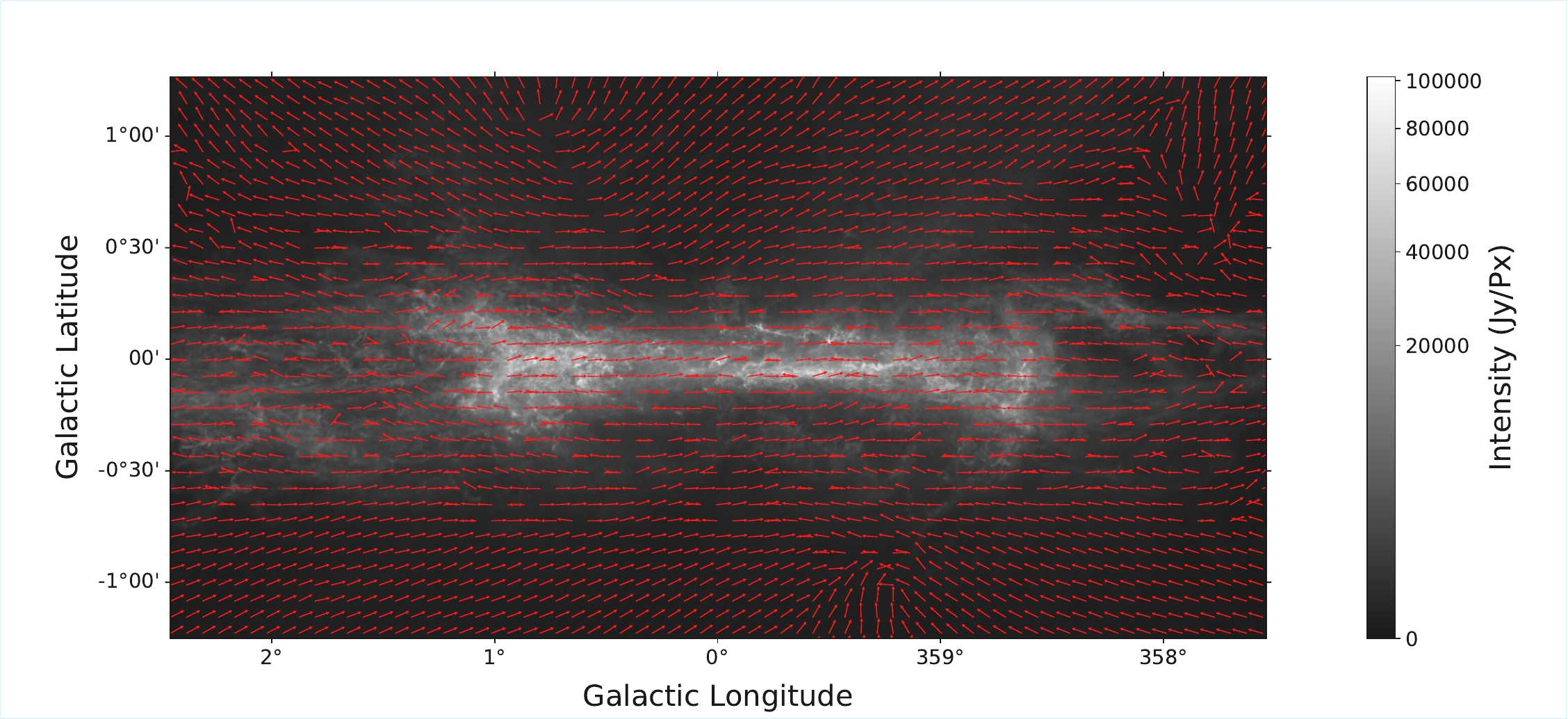}
    \caption{The modeled HNCO 4(0,4) – 3(0,3) spectral line moment 0 emission generated from \textsc{polaris}. The magnetic field vectors shown were calculated from the Q and U Stokes parameter emission maps at 214~$\mu$m and are mass-weighted along the lines of sight using the total gas density.}
    \label{fig:polaris-hnco}
\end{figure*}

\begin{deluxetable}{lcc}
\tablecaption{Synthetic Data Properties
\label{tab:physical_properties}}
\tablewidth{0pt}
\tablehead{
\colhead{Property} & \colhead{Min. Value} & \colhead{Max Value}}
\startdata
Dust Temperature (K)   & 6.2 & 15.2    \\
Dust Density (cm$^{-3}$)   & $8.2\times10^{-6}$ & $3.0\times10^{3}$ \\
Gas Temperature (K)  & 8.8 & $4.9\times10^6$    \\
Gas Density(cm$^{-3}$) & $5.3\times10^{-4}$& $3.0\times10^{5}$    \\
Velocity (km/s)   & $6.3\times10^3$ & $3.1\times10^5$     \\
Magnetic Field ($\mu$G)   & 0.1 & 300  \\
\hline
\enddata
\tablecomments{For each row the first column indicates the name of the physical property and the unit of measure in parentheses, the second row indicates the minimum value for the property, and the third column indicates the maximum value for that property.}
\end{deluxetable}

We use the radiative transfer software \textsc{polaris}, specifically designed for MHD analysis and synthetic observations, to post-process the MHD model \citep{Reissl_2016,Brauer2017,Reissl2019}. \textsc{polaris} uses a Monte Carlo approach to calculate the polarization from thermal dust emission and ray tracing to generate molecular line emission maps. We apply the code to a cutout of the MHD simulation described in Section~\ref{sec:model}, corresponding to the central cubic region with a length of 350 pc. This choice ensures that the entire CMZ region is included with minimal line-of-sight contamination from the bar dust lanes. The grid is oriented so that the observer is placed at the position of the Sun, taking into account the orientation of the dust lanes and a distance of 8.2 kpc from the center of the galaxy. We use the simulation snapshot to extract a list of parameters for each of its grid cells, such as coordinates, dust temperature, gas number density, gas velocity ($V_x$, $V_y$, and $V_z$), and magnetic field strength ($B_x$, $B_y$, and $B_z$). We calculate the gas temperature from the gas number density and the internal energy per cell provided by the simulation, assuming a helium-to-hydrogen abundance of 0.01. Similarly, we calculate the dust number density assuming it is 1\% of the gas number density. This information is converted into the \textsc{polaris}-readable input grid. 

The ranges of parameter values we obtain for all grid cells are shown in Table~\ref{tab:physical_properties}. The simulation dust temperature is comparable to that of cool CMZ clouds inferred from Herschel observations \citep[$\sim20~\rm K$;][]{Etxaluze2011,Longmore2012,Ginsburg2016,Kauffmann2017a}. The observed gas density of the CMZ is around $10^3-10^4~\rm cm^{-3}$ \citep[e.g.][]{mills2017milkywayscentralmolecular,Colzi2024}, similar to the maximum values in the simulation seen in Table \ref{tab:physical_properties}. The simulations have higher temperatures of ~$10^6$ K and velocities of thousands of \kms\ in less dense areas, while the majority of temperatures and velocities in the main dust lanes are similar to the observed CMZ values of $\sim10^2$ K \citep{Ginsburg2016,Colzi2024} and $\sim$10 \kms. The temperatures at densities comparable to those local to the HNCO filaments correspond to more reasonable temperatures of $\sim$10 K comparable to cool CMZ clouds. The magnetic field strength in the simulation is a few 100 $\mu$G in the higher density regime representative for the HNCO LFs, corresponding to the high magnetic field strengths generally inferred for the CMZ \citep{Pillai2015,Mangilli2019,Heywood2022,Lu2024,Pare2025}.

We use \textsc{polaris} to calculate the magnetic field orientation and a synthetic line emission map of HNCO. We first run the \textsc{polaris} grain alignment with the magnetic field mode to get polarization emission maps of the Stokes parameters I, Q, U, and V. \textsc{polaris} uses radiative torque (RAT) alignment to simulate the dust grains aligning with the magnetic field orientation due to the torques induced by interactions with photons in the radiation field. The dust composition is assumed to be made up entirely of silicate grains with an oblate shape and radii ranging from $5\times10^{-9}$ to $2\times10^{-6}$ m.  The electric field polarization angle is calculated at each pixel from the Q and U emission maps, and it is then rotated by 90 degrees to create the magnetic field orientation.

To generate line emission maps, \textsc{polaris} uses molecular data from the Leiden Atomic Molecular Database \citep{schoier2005atomic}. We use the HNCO 4(0,4) – 3(0,3) spectral line for the synthetic emission maps in order to match the ACES observations. We assume a constant abundance for HNCO for the whole region of $3\times{}10^{-8}$ with respect to the total gas density, which is consistent with observational measurements for the CMZ \citep{Riquelme2018}. When running the line emission radiative transfer, we make use of the large velocity gradient (LVG) approximation \citep{sobolev1957diffusion} due to the significant velocity changes throughout the molecular clouds in the CMZ. This allows radiative tranfser calculations to be done over shorter distances, as photons don't interact with gas in distant cells and escape the grid. The final HNCO moment 0 emission map is integrated over 101 velocity channels ranging from -400 \kms\ to 400 \kms, with a velocity resolution of 8 \kms. This is a lower velocity resolution than the 0.21 \kms\ resolution of the ACES band 3 observations. However, since the LFs studied in the observations are coherent velocity structures with large velocity dispersions, filamentary structures found in the synthetic data can still be compared to the observed LFs.

Figure \ref{fig:polaris-hnco} shows the final results of the synthetic data pipeline. The synthetic magnetic field orientation at 214~$\mu$m is plotted over the synthetic HNCO moment 0 line emission, and these magnetic field lines were estimated using a mass-weighted average along the lines of sight using the total gas density.

\section{RESULTS} \label{sec:res}

\begin{deluxetable*}{lccccc}
\tablecaption{Observed LF Magnetic Field Properties
\label{tab:bfield_prop}}
\tablewidth{0pt}
\tablehead{
\colhead{ID} & \colhead{$\rm HRO^{obs}_{214 \& 850}$} & \colhead{$\rm AM^{HAWC+}_{obs}$} & \colhead{$\rm AM^{POL2}_{obs}$} & \colhead{Dominant Mechanism}
}
\startdata
Region 1 & & & & & \\
LF 1a & $\perp$ & -0.48 -- -0.32 & -- & Supersonic Turbulence / Shock Compression \\
LF 1b & $\perp$ & -0.69 -- 0.05 & -- & Supersonic Turbulence / Shock Compression \\
\hline
Region 2 & & & & & \\
LF 2a & $\parallel$ & 0.30 -- 0.53 & -0.17 -- 0.16 & Subsonic Turbulence / Shear \\
LF 2b & $\parallel$ & 0.72 -- 0.89 & 0.35 -- 0.95 & Subsonic Turbulence / Shear \\
\hline
Region 3 & &  & & \\
LF 3 & $\perp$ & -0.49 -- 0.01  & -- & Supersonic Turbulence / Shock Compression \\
\hline
Region 4 & & & & \\
LF 4a & $\parallel$ to $\perp$ & -0.32 -- 0.37 & 0.38 -- 0.60 & Supersonic Turbulence / Shock Compression \\
LF 4b & $\parallel$ & 0.52 -- 0.71 & 0.31 -- 0.65 & Subsonic Turbulence / Shear \\
\hline
Region 5 & & & & & \\
LF 5 & $\perp$ & -0.68 -- -0.03 &  -0.47 -- 0.37 & Supersonic Turbulence / Shock Compression \\
\hline
Region 6 & &  & & \\
LF 6a & $\parallel$ & 0.50 -- 0.64  & -- & Subsonic Turbulence / Shear \\
LF 6b & $\perp$ & -0.83 -- -0.48 &  --  & Supersonic Turbulence / Shock Compression \\
\hline
Region 7 & &  & & \\
LF 7a & $\parallel$ & 0.07 -- 0.79 & -- & Subsonic Turbulence / Shear \\
LF 7b & $\perp$ & -0.62 -- 0.09 & -- & Supersonic Turbulence / Shock Compression\\
\hline
\enddata
\tablecomments{For each row the first column indicates the region (with regions shown in Figure \ref{fig:fil_map}) and the LFs within that region (as indicated in Figures \ref{fig:hro_rht_sofia_fig1} and \ref{fig:hro_rht_sofia_fig2}). The histogram of relative orientation ($\mathrm{HRO^{obs}_{214\& 850}}$) from the observed FIREPLACE and BISTRO observations is shown in the second column, where ``$\parallel$'' indicates that the magnetic fields are parallel with the LFs, ``$\perp$'' denotes that the magnetic fields are perpendicular to the LFs, ``$\parallel$ to $\perp$'' represents that the relative orientation between magnetic fields and LFs transits from parallel to perpendicular. The alignment measure (AM) from the observed FIREPLACE observations is shown in the third column, and the AM from the observed BISTRO observations is shown in the fourth column. The fifth column indicates the dominant mechanism likely dictating LF formation.}
\end{deluxetable*}

\subsection{Observational Results} \label{sec:obs-res}

\subsubsection{HAWC+ 214\,$\mu$m}
\begin{figure*}
    \centering
    \includegraphics[width=0.45\linewidth]{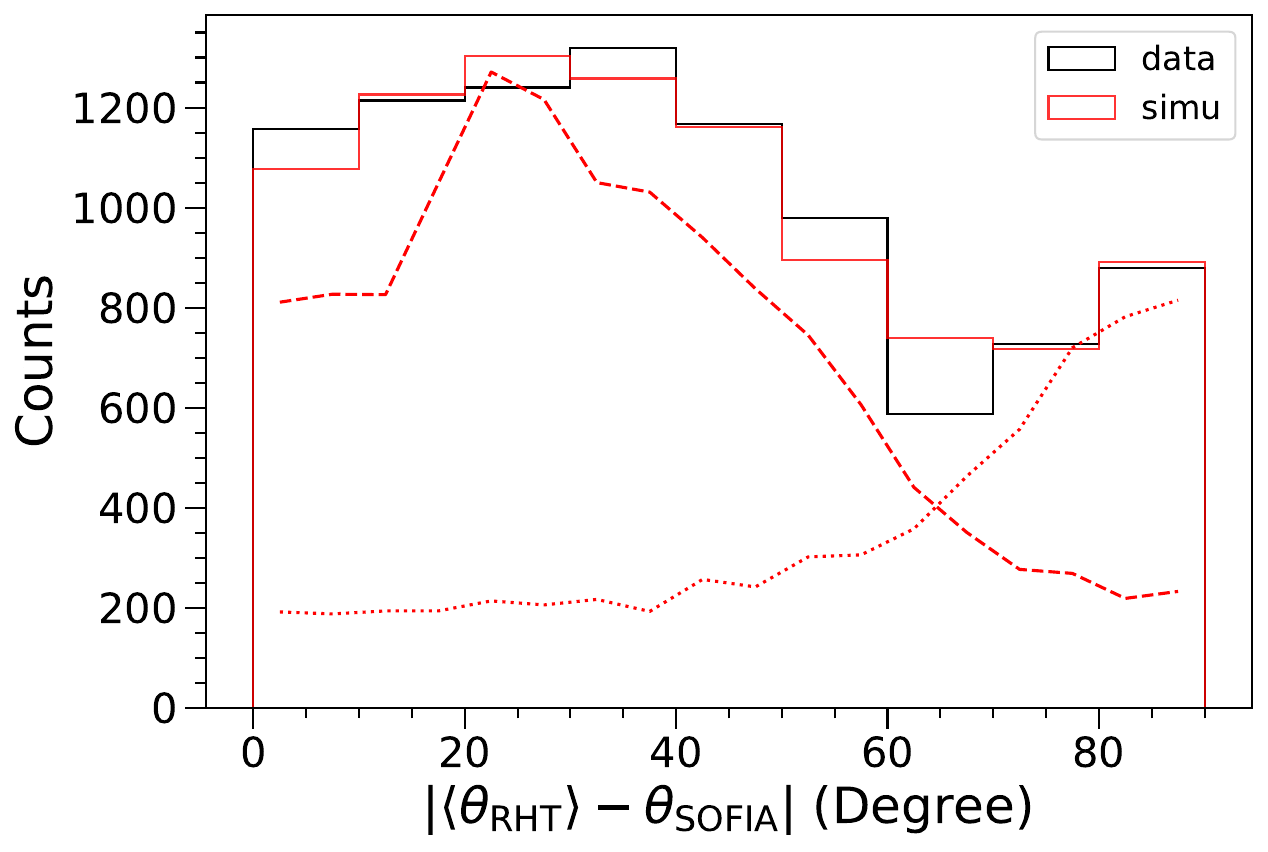}
    \includegraphics[width=0.5\linewidth]{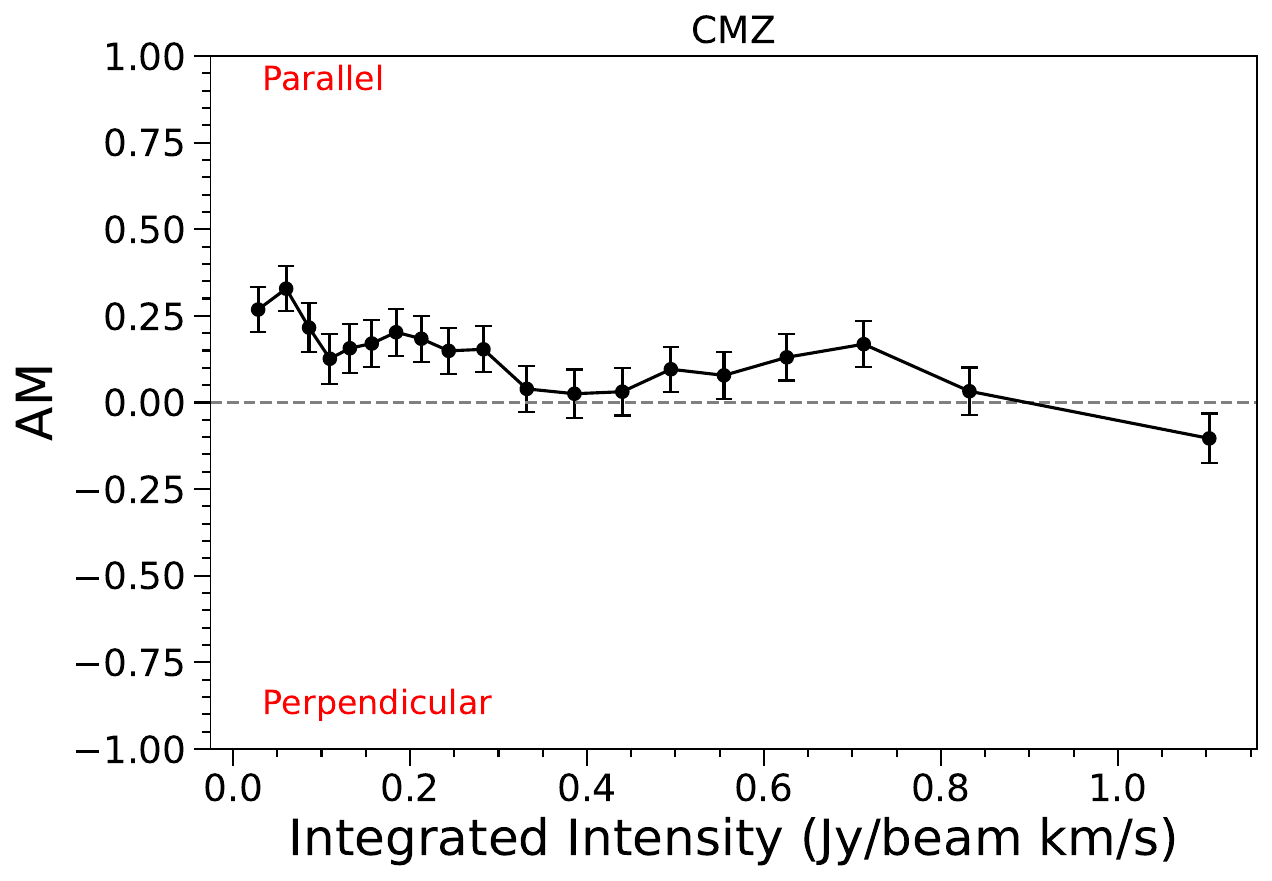} \\ 
    \caption{The relative orientation (left panel) and alignment measure (right panel) distributions obtained for the full set of targeted LFs studied using the 214~$\mu$m FIREPLACE observations. In the left panel, dashed and dotted lines show projected relative angles for intrinsic 3D angles of 20\degree to 60\degree and 75\degree to 90\degree, respectively. The red histogram is a 7:3 mix of these distributions, matching the bimodal pattern in the data.}
    \label{fig:RO_dist_SOFIA_RHT}
\end{figure*}

We find a bimodal distribution of relative angles $\phi$ (Equation \ref{eq:phi}) between the orientations of the LFs and the FIREPLACE magnetic fields, as shown in the left panel of Figure \ref{fig:RO_dist_SOFIA_RHT}. It exhibits one peak near 30\degree, indicating filaments that are roughly parallel to the magnetic field, and another peak near 90\degree, indicating filaments that are approximately perpendicular to the field. This bimodal distribution is consistent with other observational studies working on the comparisons between filaments and local magnetic field \citep[e.g.,][]{Sugitani2011,Li2013}. Observational studies \citep[e.g.,][]{Zhang2014, PlanckXXXV} have shown that projection effects can significantly influence 2D relative orientation analyses. To examine whether the bimodal distribution of $\phi$ observed in the CMZ is intrinsic, we simulate pairs of randomly oriented vectors uniformly distributed in 3D space. We select all pairs with 3D relative angles of 20\degree to 60\degree and 75\degree to 90\degree. The distribution of their projected angles are shown in the left panel of Figure \ref{fig:RO_dist_SOFIA_RHT}. It is clear that a single simulated distribution cannot reproduce the observed pattern. However, combining the two simulated distributions in a 7:3 ratio (red histogram in Figure \ref{fig:RO_dist_SOFIA_RHT}) yields a shape that closely matches the observations. A Kolmogorov–Smirnov (K–S) test comparing the combined simulation with the data gives a p-value of 0.16. This value does not provide strong evidence to reject the null hypothesis that the two distributions come from the same population, supporting the interpretation that the observed distribution is bimodal, with peaks near 30\degree and 90\degree.

To investigate how the relative orientation changes with HNCO intensities, we apply the HRO analysis to all identified LFs. The right panel of Figure \ref{fig:RO_dist_SOFIA_RHT} presents the AM (as defined in Equation \ref{eq:am}) as a function of the integrated HNCO intensity. Following \citet{PlanckXXXV2016}, we use an equal number of data points per bin to ensure consistent statistical analysis across intensity bins. Each intensity bin contains 480 pixels (corresponding to approximately 30 independent measurements), which balances the need for reliable statistics with sufficient resolution to capture trends in high-intensity regions. We confirm that varying the number of independent measurements per bin by a factor of two does not significantly alter the observed trend between relative orientation and intensities. Here, we find a decreasing trend of AM with increasing intensity, indicating that in regions with faint HNCO emission, magnetic fields tend to align with the LFs. In contrast, at regions showing strong HNCO emission, magnetic fields tend to exhibit no preferential orientation with AM values ranging from -0.25 -- 0.25.

\begin{figure*}
    \centering
    \includegraphics[width=0.9\linewidth]{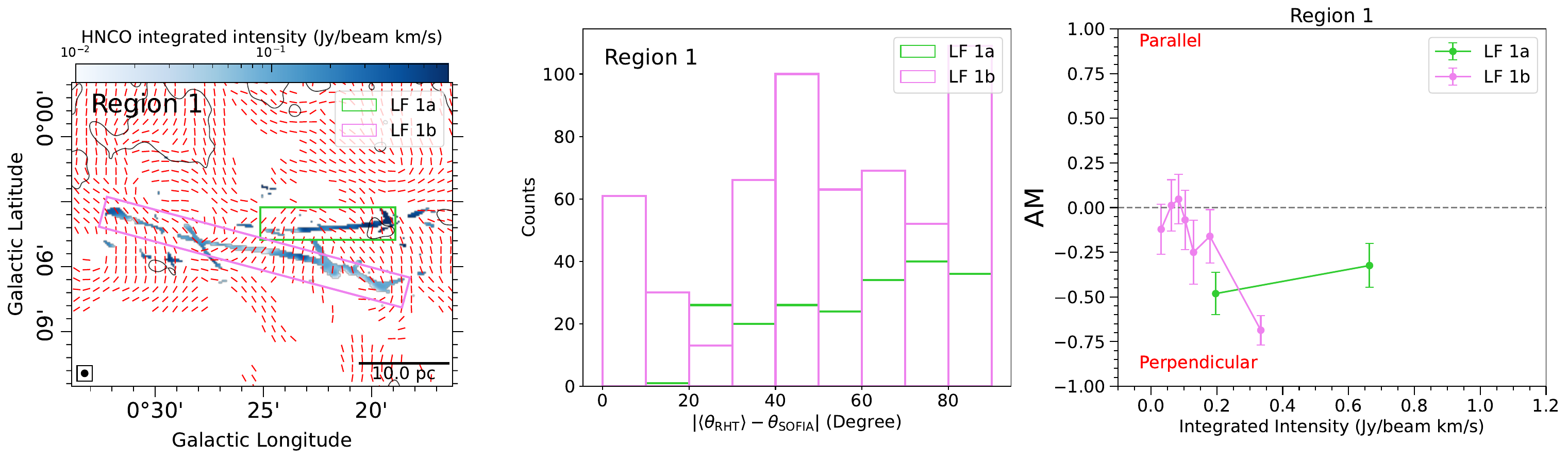}\\
    \includegraphics[width=0.9\linewidth]{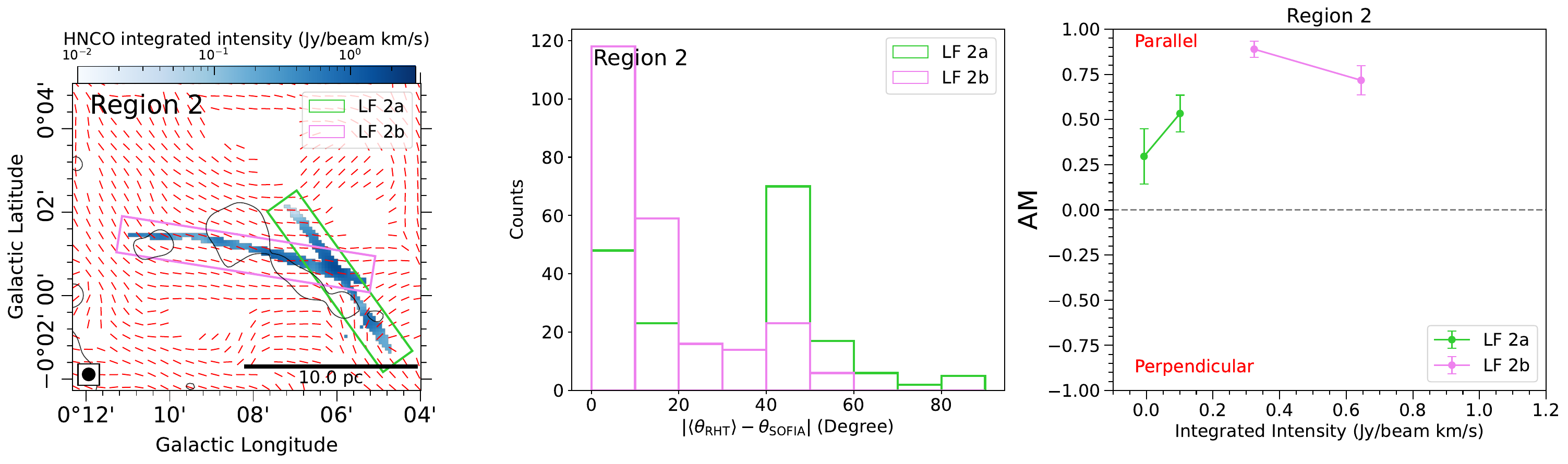}\\
    \includegraphics[width=0.9\linewidth]{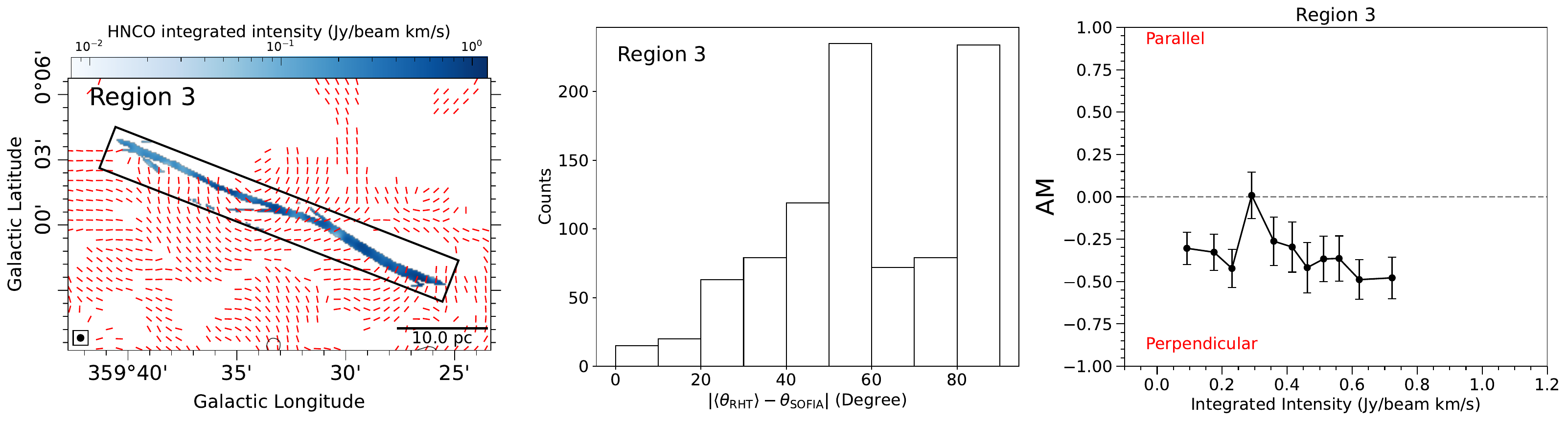}\\
    \includegraphics[width=0.9\linewidth]{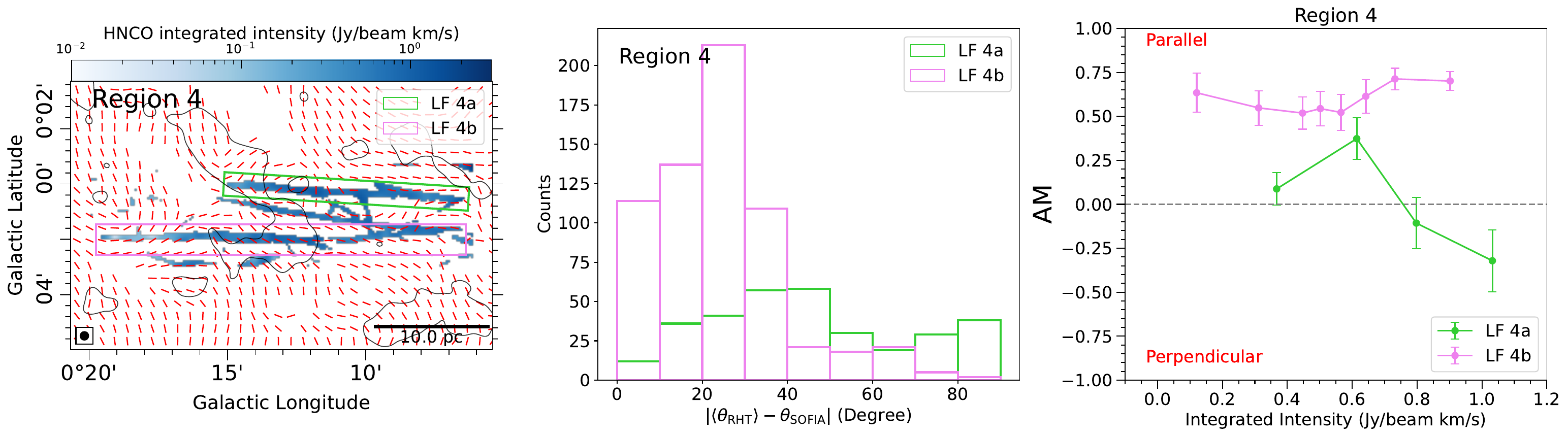}\\
    \caption{Observed relative orientation and alignment measure (AM) results obtained for individual LFs compared to the FIREPLACE (214 $\mu$m) magnetic field in the CMZ located in regions 1 -- 4. Left column: the HNCO intensity of the LFs with the non-filamentary emission surrounding the LFs masked out. Red dashes indicate the FIREPLACE magnetic field orientations. Middle column: the histogram or relative orientations between the FIREPLACE magnetic field orientation and the LF orientation. Right column: the AM measure results obtained for each LF studied.}
    \label{fig:hro_rht_sofia_fig1}
\end{figure*}

\begin{figure*}
    \centering
    \includegraphics[width=0.9\linewidth]{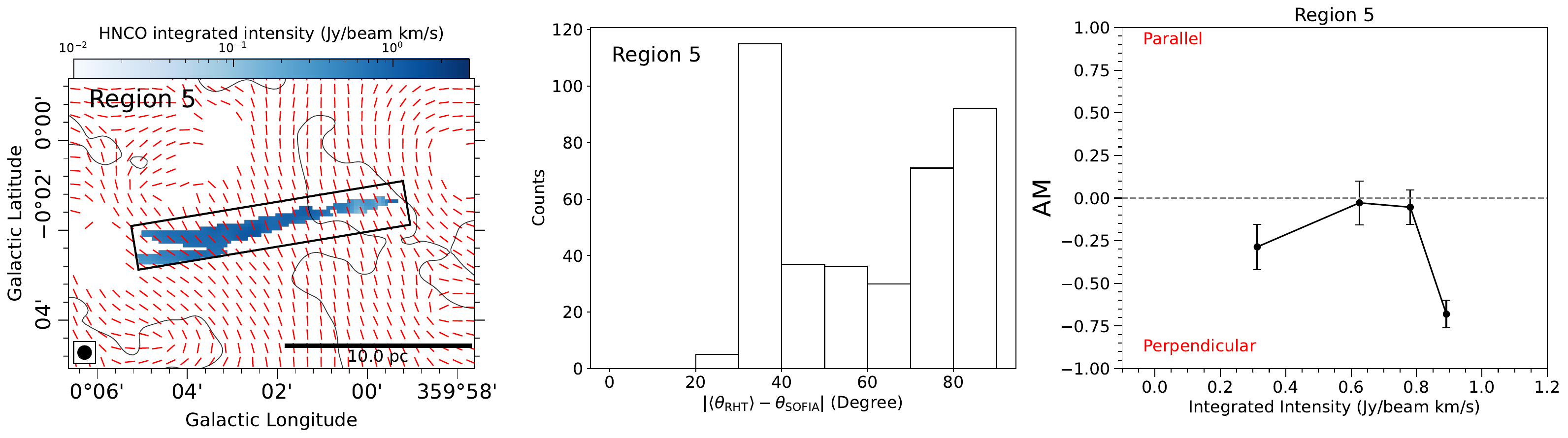}\\
    \includegraphics[width=0.9\linewidth]{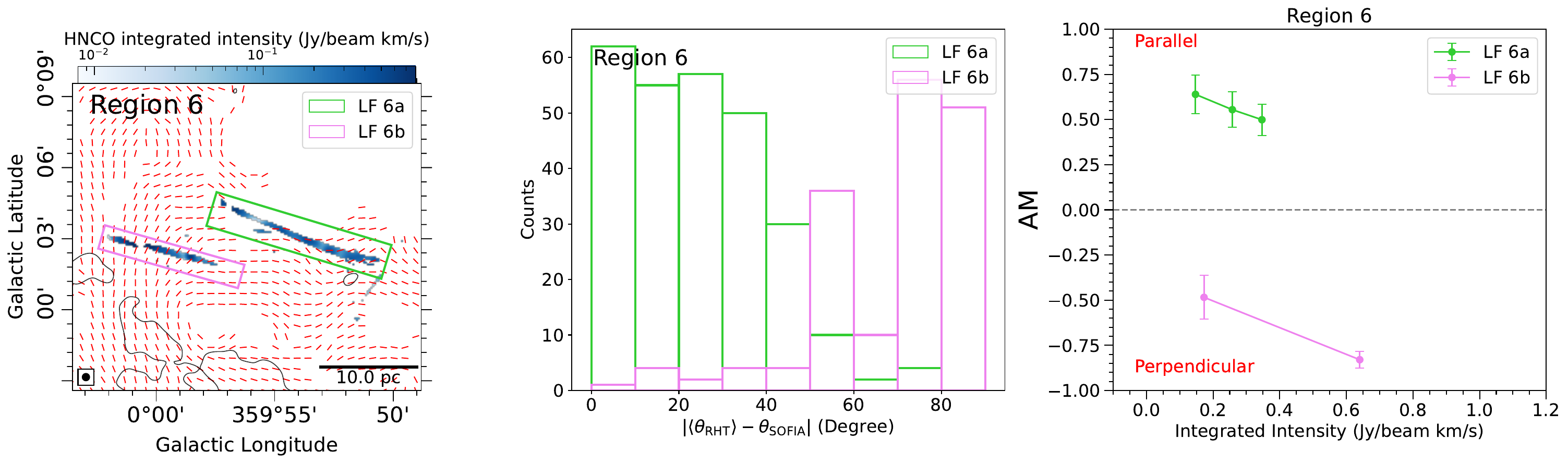}\\
    \includegraphics[width=0.9\linewidth]{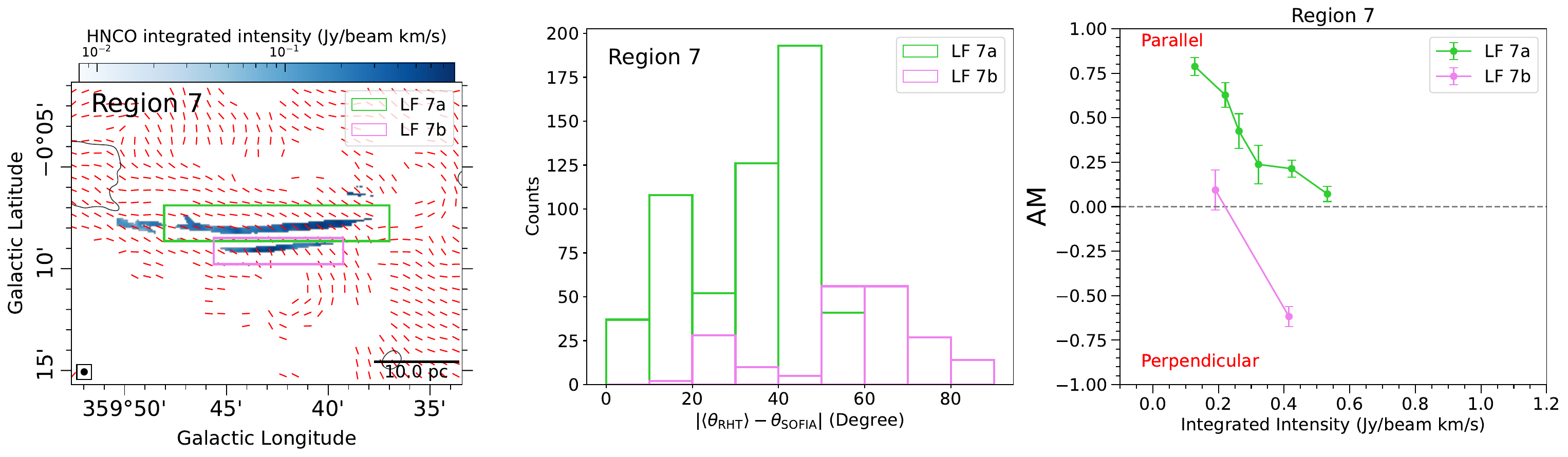}\\
    \caption{Same as for Figure \ref{fig:hro_rht_sofia_fig1} but for regions 5 -- 7.}
    \label{fig:hro_rht_sofia_fig2}
\end{figure*}

Some LFs are located near the molecular streams in the CMZ that trace out the figure eight pattern of the molecular clouds. To examine the relative orientations between the magnetic fields and the LFs in these regions, we select seven areas that lie along or cross the streams, each containing one or more LFs. In Figures \ref{fig:hro_rht_sofia_fig1} and \ref{fig:hro_rht_sofia_fig2} we show the relative orientation between SOFIA/HAWC+ magnetic fields and the LFs across the seven selected regions shown in Figure \ref{fig:fil_map}. Regions 3 and 5 show only one LF, while we identify two LFs within all the other regions. In Region 2, LFs 2a and 2b intersect on the plane of the sky but exhibit a velocity difference of over 80 $\mathrm{kms^{-1}}$ (see Table \ref{tab:hnco_prop}), suggesting that they may be unrelated components along the line of sight. Therefore, we separate them using the kinematic information. We integrate the HNCO data over the velocity ranges of -40-20 $\mathrm{km~s^{-1}}$ for LF 2a and 55-95 $\mathrm{km~s^{-1}}$ for LF 2b and derive the evolution of AM based on the integrated intensity for these different velocity ranges. It is interesting to note that even though in some regions (e.g., Regions 6 and 7), the two LFs are close to each other, they show distinct distributions of relative orientations. For example, LF 6a tends to be aligned with the magnetic field while LF 6b tends to be perpendicular to the magnetic field. The second column of Table \ref{tab:bfield_prop} summarizes the relative orientation for each LF with the third column indicating the range of AM values obtained from the FIREPLACE observations.

\subsubsection{BISTRO POL-2 850\,$\mu$m}

\begin{figure*}
    \centering
    \includegraphics[width=0.9\linewidth]{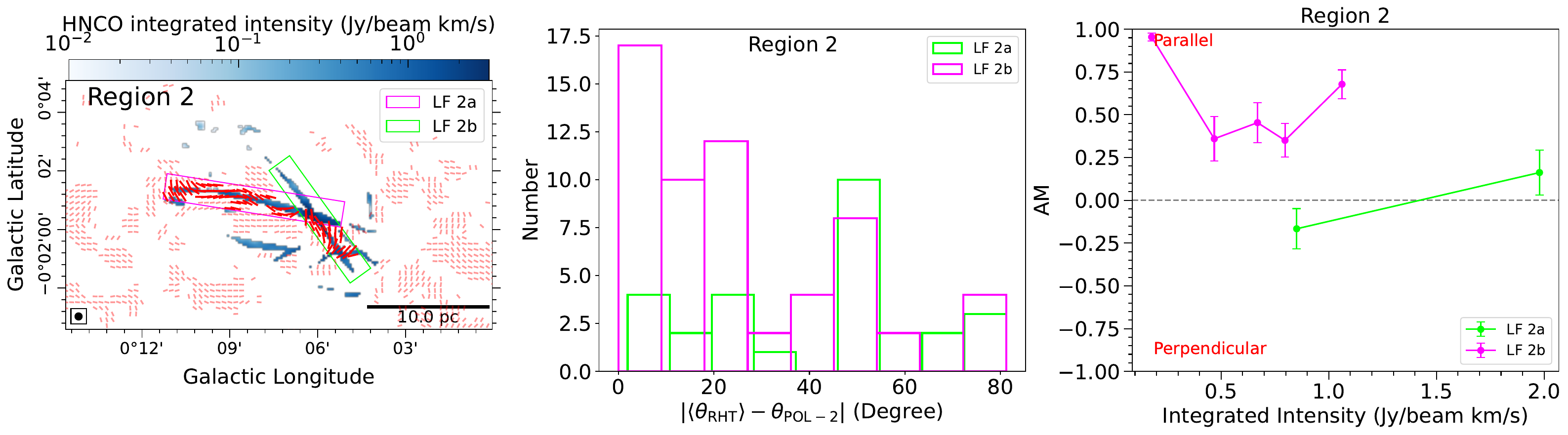}\\
    \includegraphics[width=0.9\linewidth]{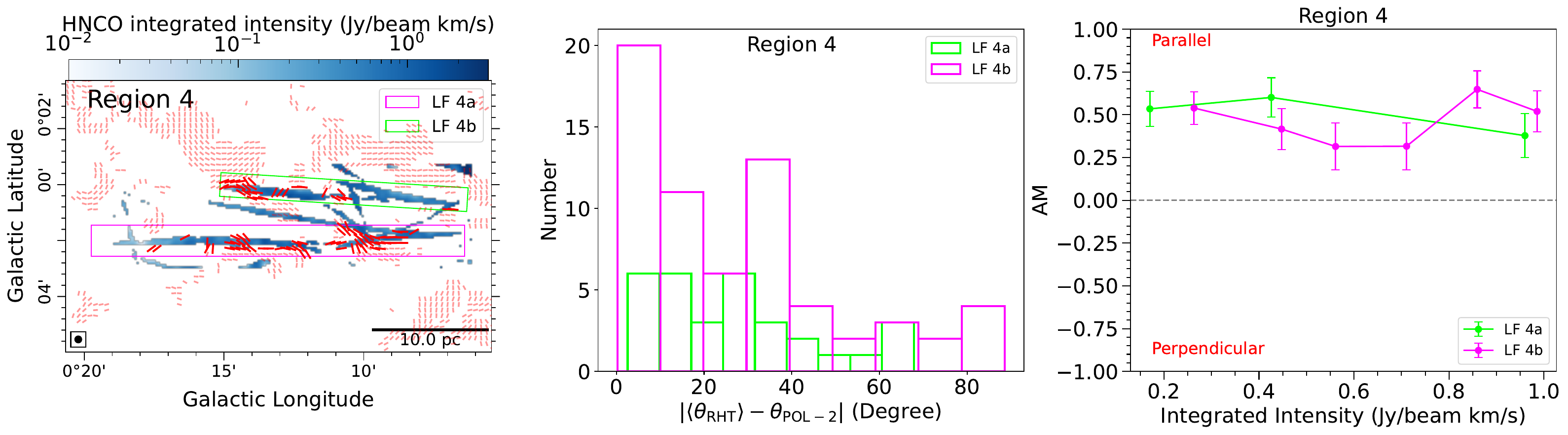}\\
    \includegraphics[width=0.9\linewidth]{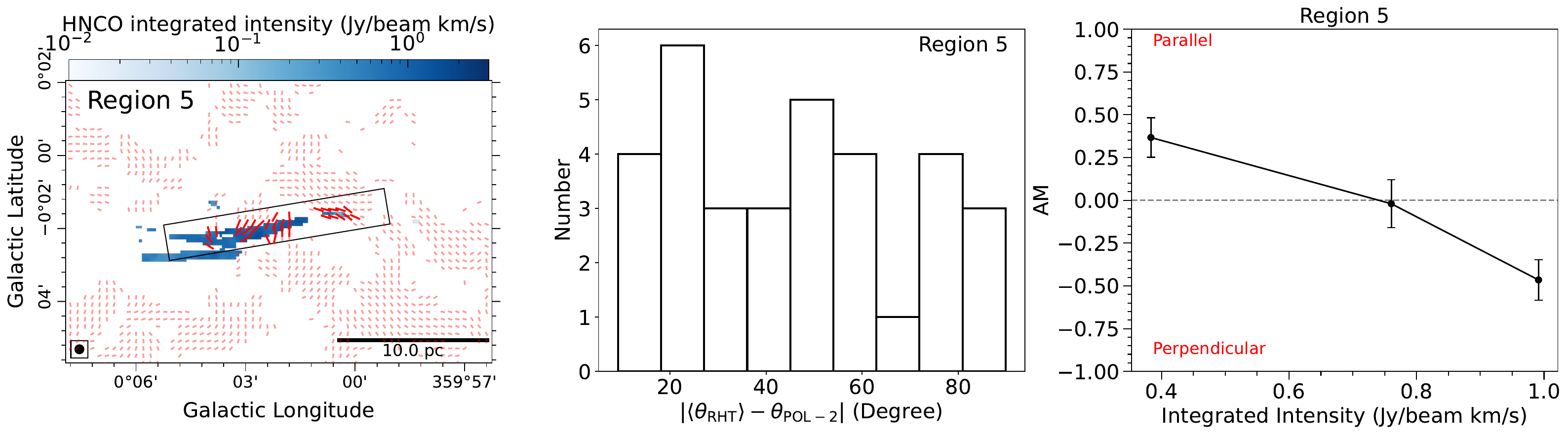}\\
    \caption{Observed relative orientation and alignment measure (AM) results obtained for individual LFs compared to the 12$\arcsec$ BISTRO (850 $\mu$m) magnetic field in the CMZ located in regions 2, 4, and 5. Left column: the HNCO intensity of the LFs with the non-filamentary emission surrounding the LFs masked out. Thicker red dashes indicate the BISTRO magnetic field vectors used to calculate the AM values. Middle column: the histogram or relative orientations between the BISTRO magnetic field orientation and the LF orientation. Right column: the AM measure results obtained for each LF studied.}
    \label{fig:hro_rht_jcmt}
\end{figure*}

In Figure~\ref{fig:fil_map} we show the three regions where we have good spatial coverage of the magnetic field at 850\,$\mu$m and identified LFs, with green and orange boxes. We see a predominantly parallel magnetic field alignment in all of the filaments studied with good spatial coverage of the 850 $\mu$m magnetic field. The JCMT, ground-based observations at 850\,$\mu$m preferentially traces the densest regions and does not recover extended emission well \citep{friberg16}. Much of the identified LFs, after removal of the molecular clouds, lie in regions of diffuse dust emission and, therefore, there is not sufficiently high signal-to-noise observations with POL-2. The JCMT results do not exhibit the bimodal magnetic field orientation that is seen from the SOFIA/HAWC+ results presented in Figure \ref{fig:RO_dist_SOFIA_RHT}. Rather, only the parallel magnetic field component (that corresponds to the peak at $\sim$30\degree\ in the left-hand panel of Figure \ref{fig:RO_dist_SOFIA_RHT}) remains. The LFs with JCMT orientations and their corresponding HRO and AM distributions are shown in Figure \ref{fig:hro_rht_jcmt} and the range of AM values obtained for these LFs are displayed in column four of Table \ref{tab:bfield_prop}.

\subsection{Synthetic Data Results}
\begin{figure*}
    \centering
    \includegraphics[width=0.99\linewidth]{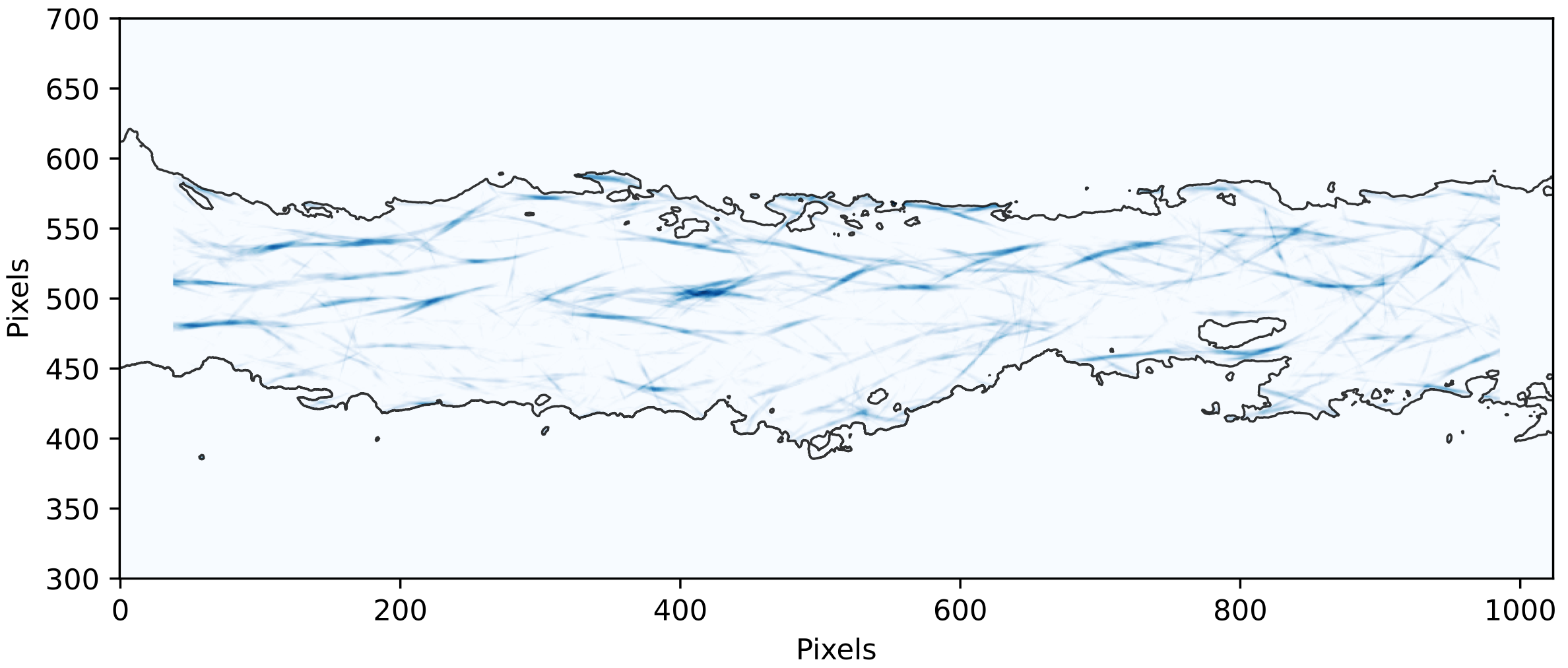}\\
    \includegraphics[width=0.99\linewidth]{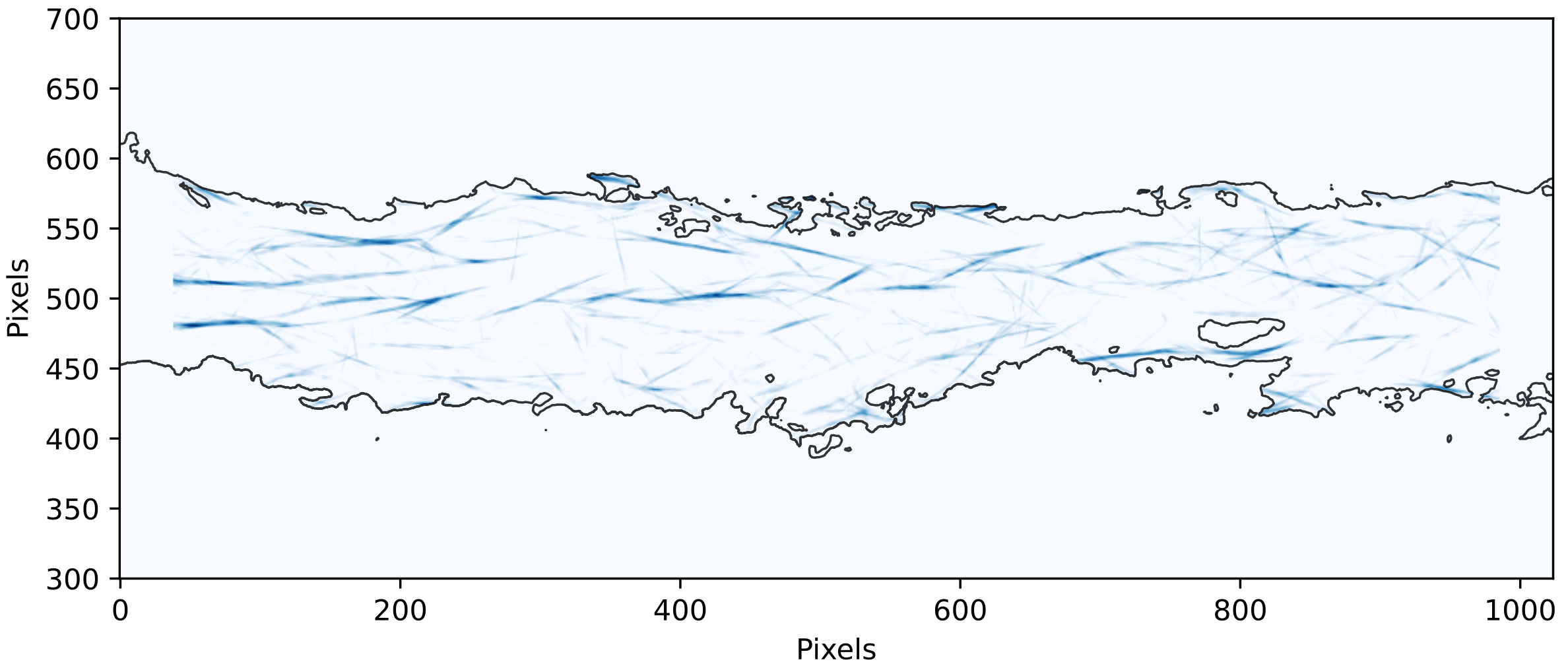}
    \caption{The RHT distributions obtained from the 214~$\mu$m (top) and 850~$\mu$m (bottom) synthetic Stokes I distributions. The contour level roughly indicates the extent of the Galactic plane with a contour value of 0.15 Jy pixel$^{-1}$.}
    \label{fig:RHT_full_synth}
\end{figure*}

\begin{figure*}
    \centering
    \includegraphics[width=0.45\linewidth]{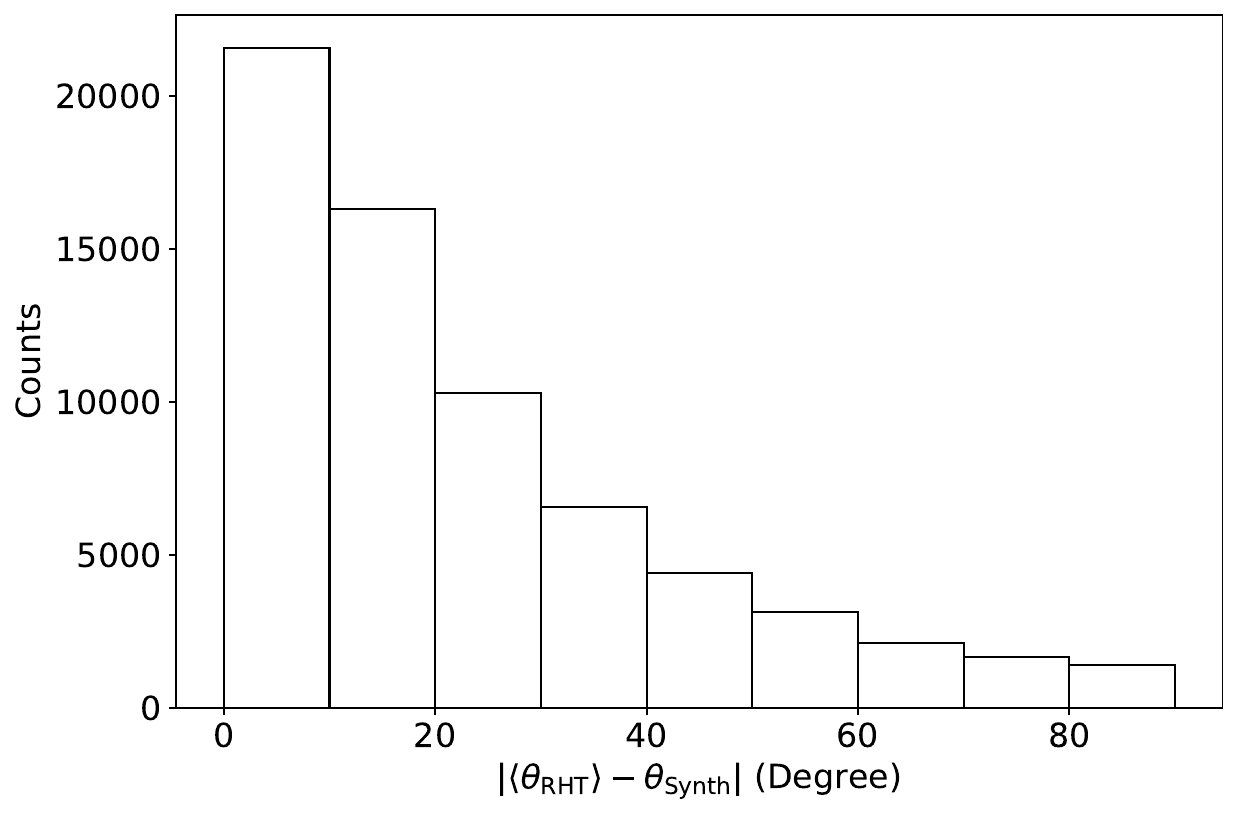}
    \includegraphics[width=0.5\linewidth]{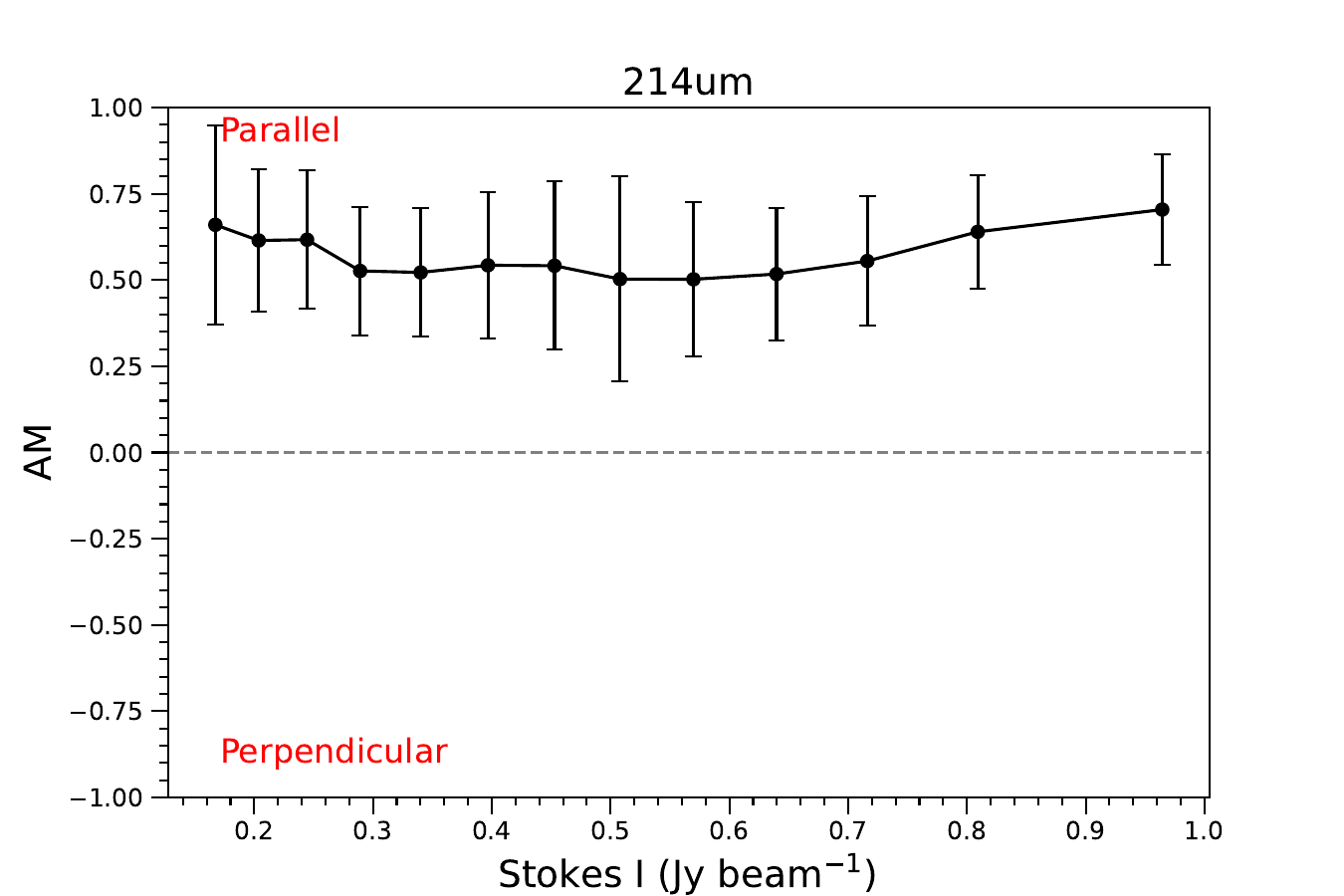} \\ 
    \includegraphics[width=0.45\linewidth]{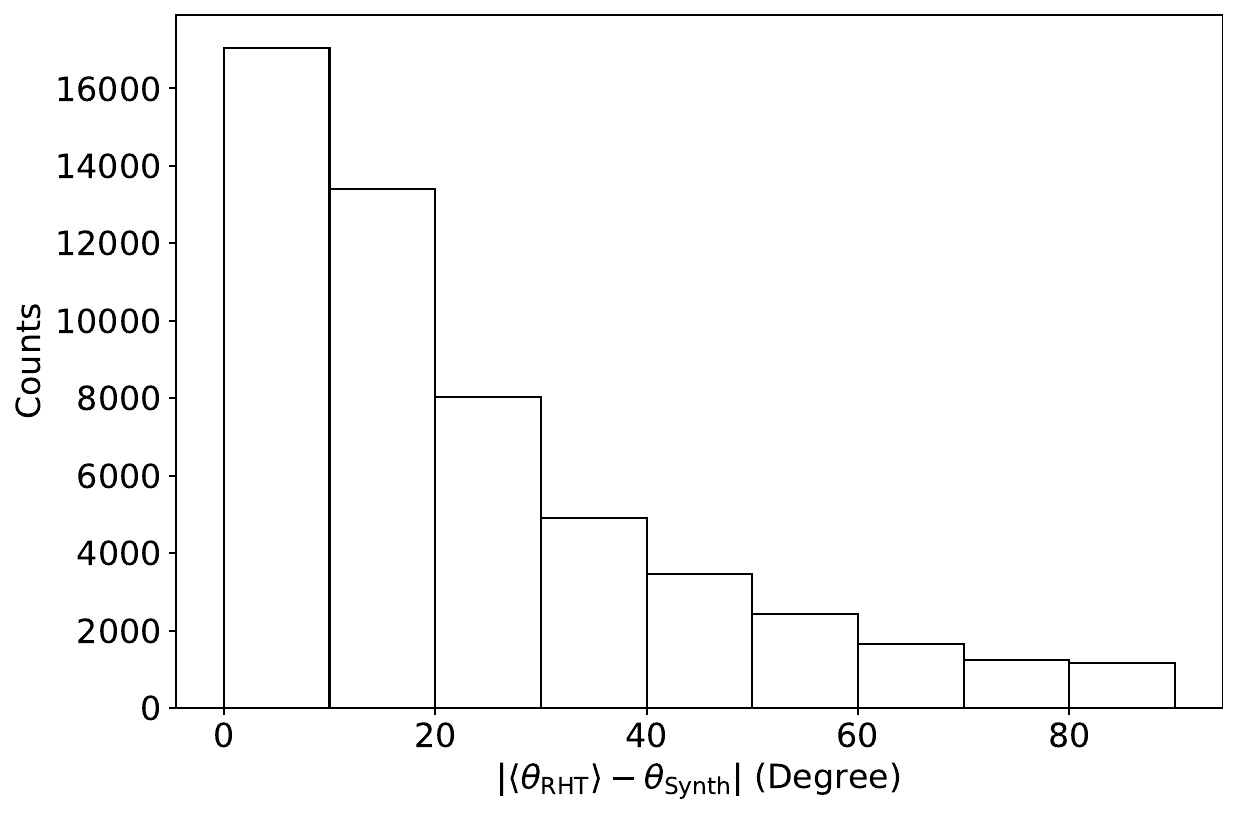}
    \includegraphics[width=0.5\linewidth]{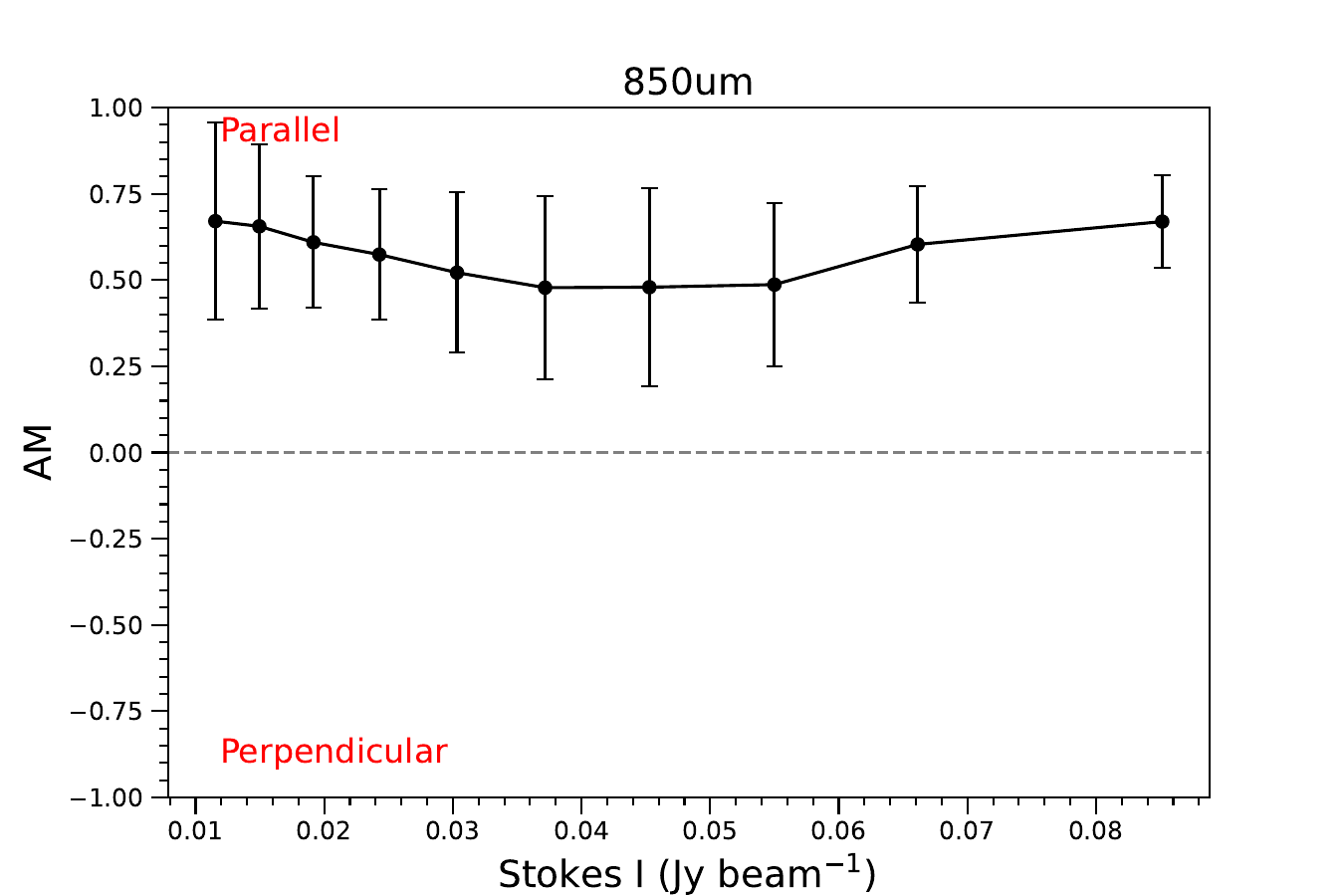} \\ 
    \caption{The relative alignment results for all identified filaments in the synthetic data sets. Left column: the distribution of relative orientations, shown as histograms, for the 214~$\mu$m (top) and 850~$\mu$m data (bottom) sets. Right column, the AM results obtained for the two wavelengths.}
    \label{fig:RO_AM_synth}
\end{figure*}

\begin{figure*}
    \centering
    \includegraphics[width=0.9\linewidth]{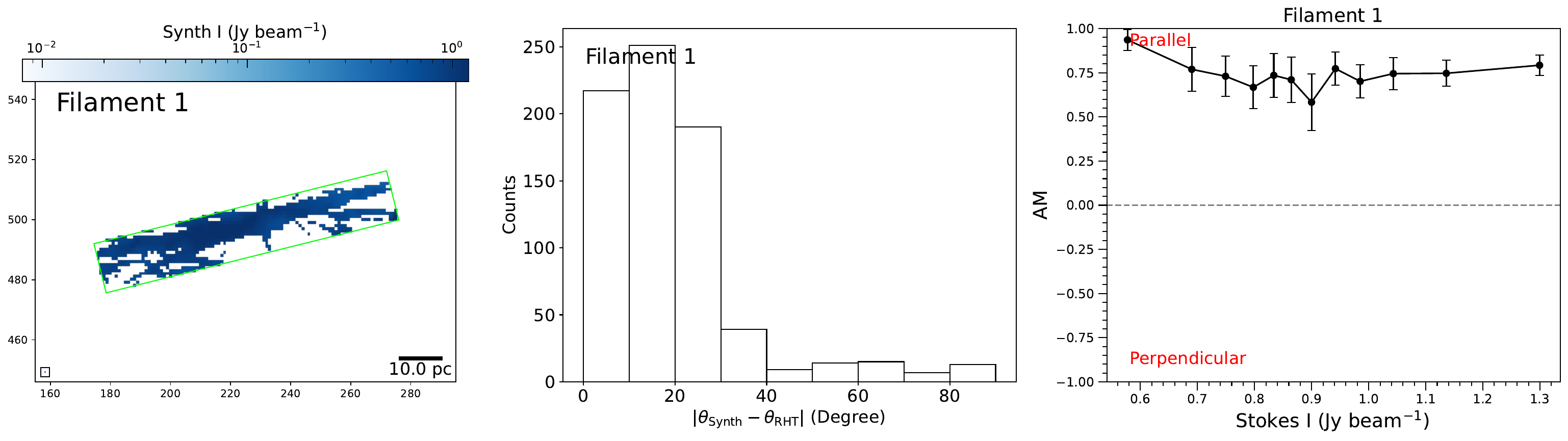}\\
    \includegraphics[width=0.9\linewidth]{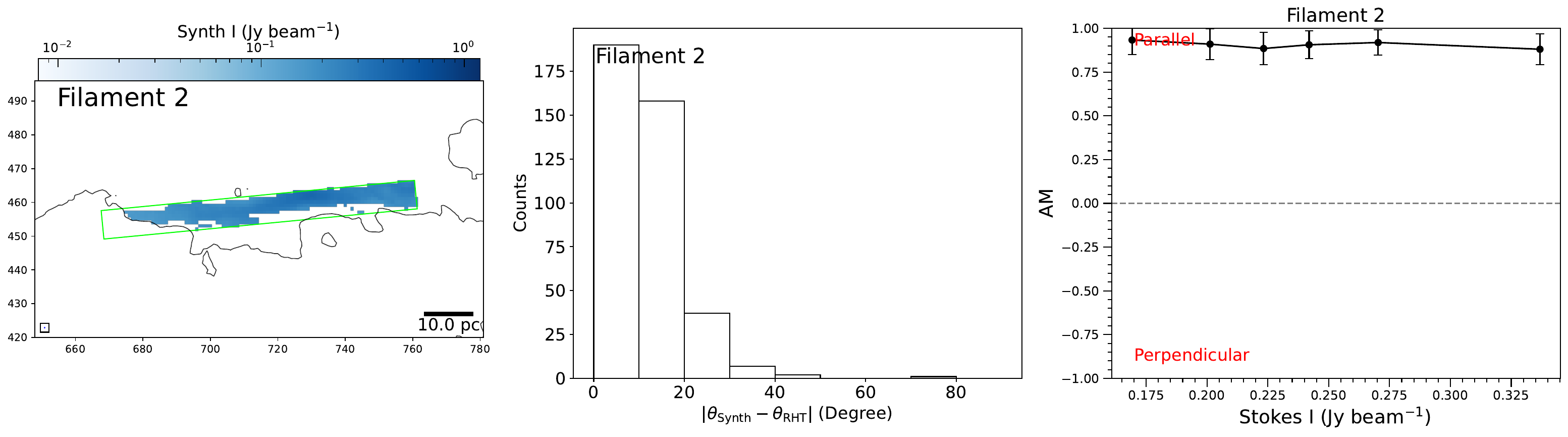}\\
    \caption{Relative orientations and alignment results for the larger filaments identified in the 214~$\mu$m synthetic data set. Left column displays the synthetic filament. The middle column shows the distribution of relative orientations as histograms. The right column shows the AM values obtained for the filaments.}
    \label{fig:filament_214_AM_RO}
\end{figure*}

\begin{figure*}
    \centering
    \includegraphics[width=0.9\linewidth]{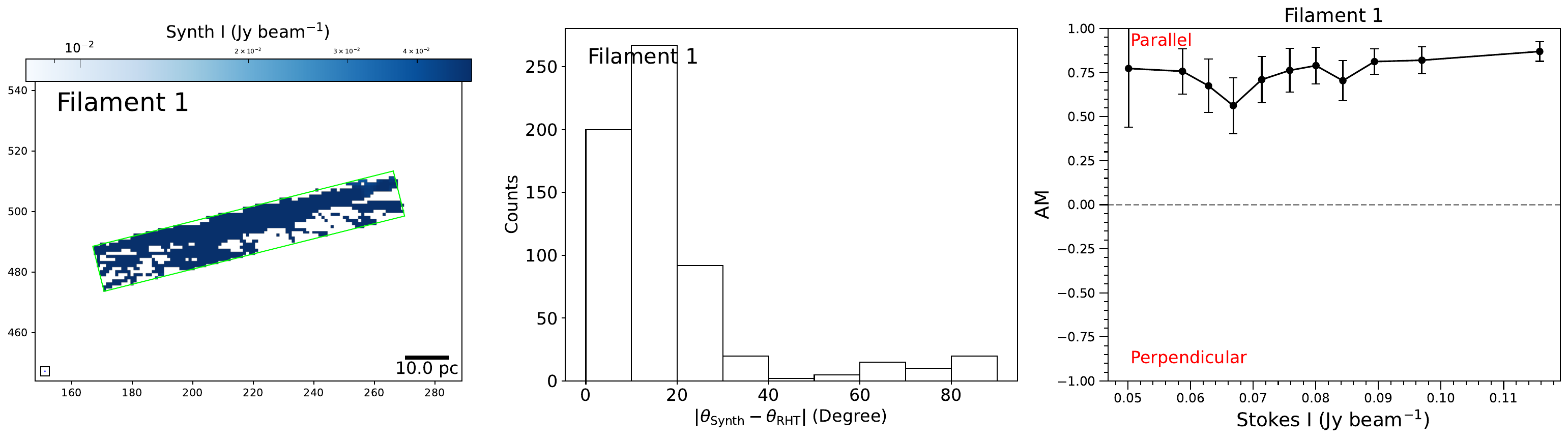}\\
    \includegraphics[width=0.9\linewidth]{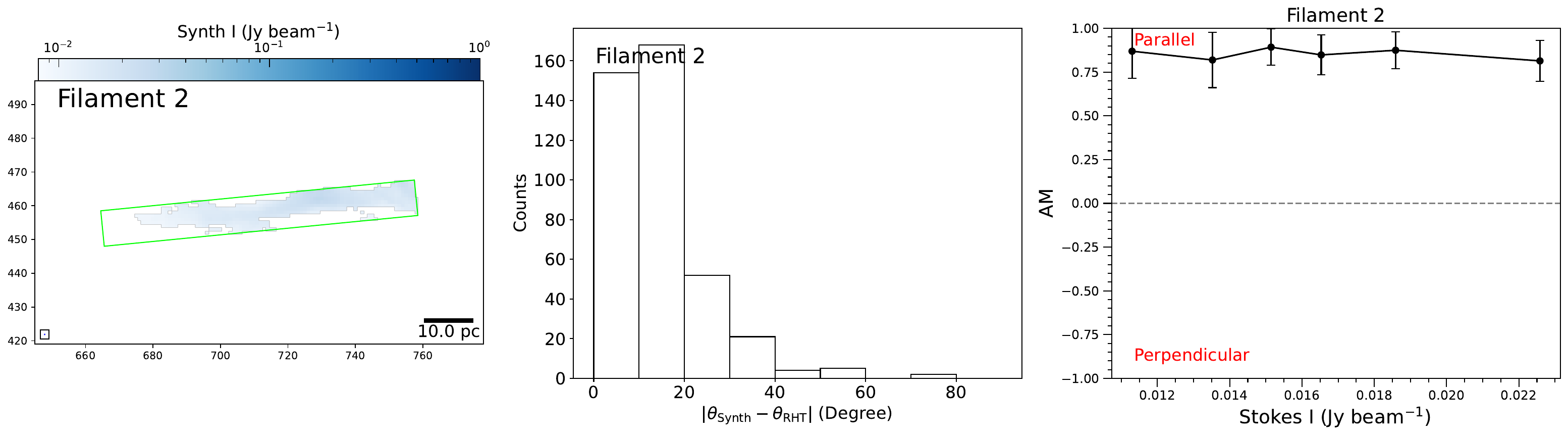}\\
    \caption{Same as for Figure \ref{fig:filament_214_AM_RO}, but for the same filaments as identified in the 850~$\mu$m synthetic data set.}
    \label{fig:filament_850_AM_RO}
\end{figure*}

\begin{deluxetable}{lcc}
\tablecaption{Synthetic Filament Magnetic Alignment Results
\label{tab:bfield_synth_prop}}
\tablewidth{0pt}
\tablehead{
\colhead{Filament} & \colhead{AM$^{214}_{Synth}$} & \colhead{AM$^{850}_{Synth}$}}
\startdata
Filament 1  & 0.58 -- 0.94 & 0.56 -- 0.87 \\
Filament 2  & 0.88 -- 0.93 & 0.81 -- 0.89 \\
\hline
\enddata
\tablecomments{For each row the first column indicates the filament identified within the larger synthetic data set. The range of AM values found for this filament from the 214~$\mu$m synthetic data is shown in the second column and for the 850~$\mu$m synthetic data in the third column.}
\end{deluxetable}

To compare to the observational results presented in Section \ref{sec:obs-res} we create two synthetic data sets, one at 214~$\mu$m and one at 850~$\mu$m. These data sets were made using the magnetic field strengths derived from the MHD model described in Section \ref{sec:model}, meaning these synthetic data sets represent the regime where the magnetic field likely dominates over other mechanisms like turbulence. We then identify HNCO filaments in these synthetic data sets by employing the same filament identification strategy described in Section \ref{sec:meth-id}. Examples of the distribution of identified filaments in the synthetic data sets are shown in Figure \ref{fig:RHT_full_synth}.

We first inspect the global relative orientation and alignment measure distributions obtained for the full RHT distributions obtained for the different wavelengths. We show these distributions in Figure \ref{fig:RO_AM_synth}. We find that at both wavelengths the magnetic field is predominantly oriented parallel to the HNCO filament orientation. This result is in agreement with what is observed by the 850~$\mu$m relative orientation results shown in the right column of Figure \ref{fig:hro_rht_jcmt}.

We isolate two filaments that are identified in both the 214 and 850~$\mu$m RHT distributions. These filaments have $\geq$10 pc lengths making them similar in morphology to the LFs studied in the HNCO observational data.  We study the relative orientation and alignment measure distributions corresponding to these individual filaments in Figure \ref{fig:filament_214_AM_RO} for the 214~$\mu$m synthetic data set and Figure \ref{fig:filament_850_AM_RO} for the 850~$\mu$m synthetic data set. We again find a predominantly parallel magnetic field orientation for both synthetic filaments at both wavelengths.

\section{DISCUSSION} \label{sec:disc}

\subsection{Characterization of Identified LFs} \label{subsec:filaproper}
Table \ref{tab:hnco_prop} indicates key properties of the LFs such as their central coordinates, lengths, widths, and representative HNCO moment 0 intensities. These LFs generally have lengths of $\sim$10 pc or longer. All of the LFs have widths of $\sim$0.5 pc, where the widths were determined through inspection of the masked HNCO moment 0 distribution that has been smoothed to the 13\arcsec\ RHT kernel size. The lengths and widths of the LFs make them comparable to the LFs studied in Battersby et al. (submitted) that were thought to be components of larger CMZ orbital structures. This thought is supported by the fact that the LFs generally trace the molecular stream structures observed in the CMZ. Furthermore, the aspect ratios of these structures ($>12$) make them comparable to the Galactic plane bones studied by FIELDMAPS and other investigators \citep[e.g.,][]{Goodman2014,Zucker2018,Stephens2022,Coude2025}.

Based on the morphological and velocity properties of the LFs studied in Battersby et al. (submitted), they concluded that these structures were likely in the CMZ rather than being foreground structures. We have performed a similar analysis in this work and again identify that the targeted LFs have broad line widths of 10s of \kms\ and are generally oriented along the larger orbital structures in the CMZ, justifying the conclusion that these are structures local to the CMZ. The tendency of the LFs to trace the orbital structures in the CMZ matches the behavior observed for the large Galactic Disk filaments that trace the spiral structure of the Galaxy \citep{Reid2014,Goodman2014}. The LFs observed in this work also align with a population of thermal, short filaments identified in MeerKAT 1 GHz observations of the CMZ \citep{Yusef-Zadeh2023}. The LFs studied here are much longer than the short filaments in \citet{Yusef-Zadeh2023}, but do appear to be generally parallel.

We argue that the LFs studied in this work are comparable to the larger filaments studied in the Galactic Disk (such as the bones studied by FIELDMAPS). The LFs studied here are all velocity coherent with aspect ratios $\geq$10. Furthermore, the LFs trace the molecular streams that are thought to connect to the Galactic bar whereas the bones trace the spiral arms of the Galactic Disk. We do note, however, that the velocity dispersions of the LFs are large, (being $\geq$10 \kms) since they are within the GC, and many of them are farther from the Galactic mid-plane than the larger filaments in the Galactic Disk. Because of the parallels between the LFs in the CMZ and the bones in the Galactic Disk we propose the use of the term ``ribs'' to refer to the CMZ LFs as an extension of the skeletal structure analogy used to define the larger filaments (``bones'') in the Galactic Disk.

\subsection{Comparing the 214 and 850~$\mu$m Magnetic Field Alignments}
\begin{figure}
    \centering
    \includegraphics[width=0.49\textwidth]{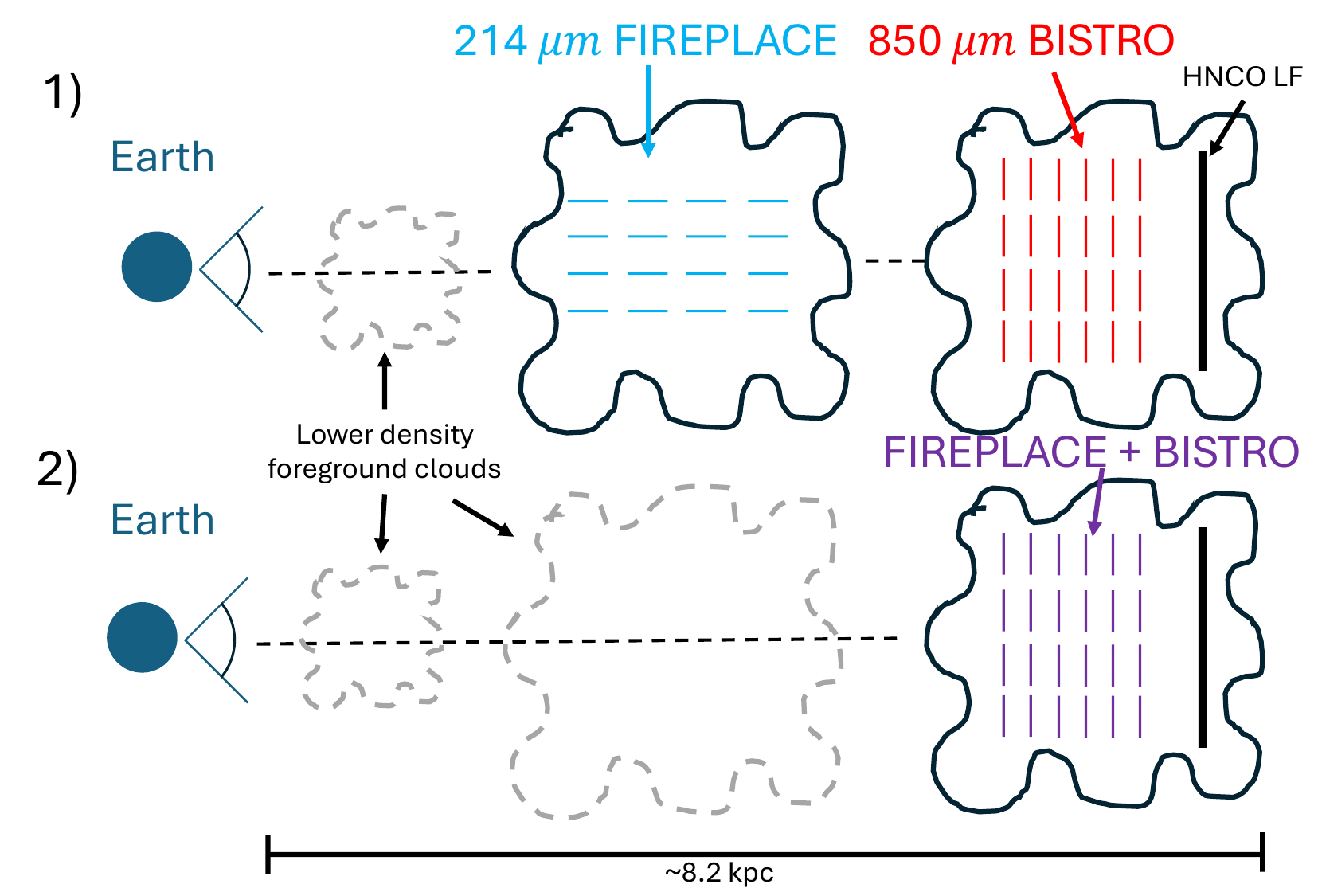}
    \caption{A sketch illustrating representative magnetic field geometry along the line of sight towards one of the GC HNCO LFs studied in this work. In the upper row (scenario 1), the FIREPLACE magnetic field traces a foreground high density medium that is not local to the HNCO LF, whereas the BISTRO magnetic field traces a high density medium local to the LF. In the lower row (scenario 2), the foreground structure that FIREPLACE was tracing in scenario 1 is lower density along this line-of-sight, and so both FIREPLACE and BISTRO trace the same high density structure that is local to the HNCO filament in this case, causing the BISTRO and FIREPLACE observations to trace the same field system.}
    \label{fig:B-sketch}
\end{figure}
The BISTRO/POL-2 observations largely coincide with the dense CMZ molecular clouds which are masked out for this work. However, several of the LFs targeted in this study do have significant BISTRO polarization orientations, specifically regions 2, 4, and 5.

The LFs in these regions show a ubiquitously parallel orientation with the magnetic field inferred from the 850~$\mu$m BISTRO observations as can be seen in Figure \ref{fig:hro_rht_jcmt}. This parallel orientation indicates that the bimodal distribution observed at 214~$\mu$m is not present at 850~$\mu$m. In particular, the 90\degree\ orientation component observed in the left-hand panel of Figure \ref{fig:RO_dist_SOFIA_RHT} is not present in the comparison to the BISTRO magnetic field presented in Figure \ref{fig:hro_rht_jcmt}.

An important note regarding the magnetic fields studied in this work is that the beam sizes of both the JCMT (12\arcsec, 0.5 pc) and FIREPLACE (19.6\arcsec, 0.8 pc) observations are larger than the widths of the LFs observed. These observations might therefore not be sensitive to the magnetic field within the filament itself, but could rather be tracing the field in the environment surrounding the filament. Follow-up observations with smaller angular resolution ($\sim$1\arcsec) are needed to probe any field variations that may be occurring immediately local to the LFs.

Nonetheless, we can make some inferences from the observations studied here. The filaments coinciding with significant BISTRO vectors are located in the center of the GC, with other filament regions studied using the 214~$\mu$m observations at the edge of the GC not covered. The 90\degree\ magnetic field component observed at 214~$\mu$m could therefore be originating from this outer regime of the GC. Alternatively, the 90\degree\ component could originate from warm dust which the 214~$\mu$m observations would be more sensitive to than the 850~$\mu$m BISTRO observations.

The differences in AM observed between the 214~$\mu$m and 850~$\mu$m magnetic fields could be a result of the complex magnetic field geometry observed towards the GC. The FIREPLACE and BISTRO data sets are most sensitive to different dust temperatures, with the 214~$\mu$m FIREPLACE observations being more sensitive to warmer dust than the 850~$\mu$m BISTRO observations. 

To help illustrate this point regarding the different magnetic field systems we present a sketch of a representative magnetic field geometry for a typical LF in Figure \ref{fig:B-sketch}. Both the FIREPLACE and BISTRO magnetic fields are density-weighted, but the BISTRO observations are possibly tracing the high-density gas that is more local to the HNCO filaments. Figure \ref{fig:B-sketch} represents this by showing the 850 $\mu$m field traced by BISTRO located in a cloud structure local to the HNCO filament. For the 214 $\mu$m FIREPLACE magnetic field the location of the high density component likely varies throughout the GC which is illustrated by the two scenarios shown in the sketch. In the first scenario (labeled 1 in Figure \ref{fig:B-sketch}), the FIREPLACE magnetic field is tracing a foreground gas structure (though this foreground structure is likely still located in the central kpc of the Galaxy), which is where the 90\degree\ field component identified in Figure \ref{fig:RO_dist_SOFIA_RHT} possibly originates from. In the lower scenario (labeled 2 in Figure \ref{fig:B-sketch}), the FIREPLACE magnetic field traces the dust local to the HNCO filament, resulting in the parallel field enhancement observed in Figure \ref{fig:RO_dist_SOFIA_RHT}.

We can also compare our results to the Galactic Disk bones studied by FIELDMAPS \citep{Stephens2022,Coude2025}. FIELDMAPS finds a generally perpendicular magnetic field for the bones, although there is evidence in some of the bones, like in  G47, that a parallel or mixed field is also present in the lower-density portions of the bones \citep{Stephens2022}. The generally perpendicular orientations seen in the FIELDMAPS bones agrees with the perpendicular field orientations exhibited by LFs 1a, 1b, 3, 5, 6b. However, the remaining LFs studied here demonstrate a parallel or mixed magnetic field orientation that parallels what is observed in the lower-density portions of the FIELDMAPS bones.

We also analyze how the magnetic field and filament orientations compare to the velocity field direction within the CMZ. Table \ref{tab:hnco_prop} displays the HNCO velocities recovered for the LFs as observed by ACES (Walker et al. 2025, submitted). From this table we can see that the filaments that are in the Galactic East of the CMZ generally have positive velocities of a few 10s of \kms, whereas those in the Western extent of the CMZ generally have negative velocities of 10s of \kms. This agrees with the CMZ-wide HNCO moment 1 distribution presented in Walker et al. (2025, submitted) where they observe positive velocities in the Eastern portion of the CMZ and negative velocities in the Western portion. We do not, however, observe a strong dependence on the magnetic field orientation with respect to the HNCO velocity. The predominant magnetic field orientation is shown in column 2 of Table \ref{tab:bfield_prop}, and the filaments with positive velocities in the Eastern extent of the CMZ (regions 1, 2, and 4) show a roughly even mixture of parallel and perpendicular magnetic field orientations. We see the same mixing of magnetic field alignment for the remaining regions that are in the Western portion of the CMZ. The major takeaway of the analysis presented here is that the different field systems observed for the LFs could indicate different physical conditions local to the different LFs that do not necessarily connect to the larger geometry of the CMZ.

\subsection{Comparison to Synthetic data}
The ubiquitously parallel AM results obtained for the synthetic filaments agree with what is observed for some observational LFs like those in Region 4 shown in Figures \ref{fig:hro_rht_sofia_fig2} and \ref{fig:hro_rht_jcmt}. However, the perpendicular and mixed field systems identified in other LFs are not reproduced by the synthetic data at either wavelength.

We note that the synthetic data set analyzed here is produced from the MHD model which is imbued with a high magnetic field strength (Table \ref{tab:physical_properties}). These large field strengths allow us to probe what the orientation of the B-field is in magnetic field dominated filaments. Such an environment is representative for the CMZ which has elevated magnetic field strengths relative to the Galactic Disk \citep{Morris1996b}.

Parallel magnetic field alignment is generally observed in LFs that are dominated by subsonic turbulence or shear (Table \ref{tab:bfield_prop}). It has been argued previously that mechanisms like subsonic turbulence and shear could inhibit cloud collapse \citep[e.g.,][]{MacLow2004}. Specific to the CMZ, shear-dominated driving of turbulence has been argued to significantly inhibit star formation for some CMZ clouds \citep{Federrath2016a,Barnes2017,Pare2025}. Our results therefore indicate that the magnetically dominated synthetic filaments studied exhibit similar AM and Histogram of Relative Orientation (HRO) distributions for LFs dominated by subsonic turbulence. We therefore argue that this finding supports the possibility that magnetic field dominated field systems are also supported against cloud collapse.

An alternative scenario we consider is that of cloud-cloud collisions. In regions where the magnetic field dominates, dense gas would form at the collision fronts of the clouds. These fronts would host the brightest HNCO and dust emission when observed edge-on. In the case of cloud-cloud collision, the largely parallel magnetic fields could indicate converging flows from above and below the Galactic plane. Converging flows like this would be expected in environments with significant stellar feedback, like the CMZ. There is ample evidence of previous starburst events and outflows from the CMZ \citep{Heywood2019,Ponti2021}, and with the steep gravitational potential of the CMZ much of this material could fall back to the Galactic plane. Such converging winds could serve as an alternative origin for the formation of the LFs.

The magnetic field alignment is not ubiquitously parallel to the HNCO filaments, however. The filaments that have a perpendicular alignment to the magnetic field could have been formed through shock compression \citep{2019ApJ...886...17H}. This shock compression could have been introduced by supersonic turbulence, facilitating cloud compression and star formation \citep{MacLow2004,Federrath2012}. The range of magnetic field alignment seen throughout the LFs is further evidence of the changing local conditions within the CMZ.

Further work using MHD models made with lower magnetic field strengths will be an important future comparison to test possible LF formation scenarios like the ones discussed here.

\section{CONCLUSIONS} \label{sec:conc}
In this work we have studied the relative orientation between a subset of large HNCO filaments (LFs) identified in the ACES observations (3.4 mm) and the magnetic field inferred from cool dust observations made using SOFIA/HAWC+ (214~$\mu$m; FIREPLACE survey) and JCMT/POL-2 (850~$\mu$m; BISTRO survey). We employ the Alignment Measure (AM) and Histogram of Relative Orientation (HRO) methods to study how the orientation of the magnetic field compares to the orientations of the identified LFs. We then compare our results to synthetic data sets made from an MHD model of the CMZ (Sections \ref{sec:model} and \ref{sec:synth}). We summarize our findings here:
\begin{itemize}
    \item We focus our analysis on the LFs that do not coincide with CMZ molecular clouds. These LFs have characteristic lengths of $\sim$10 pc and velocity magnitudes and dispersions generally $\geq$10 \kms. They are morphologically similar and are seen throughout the GC. Their properties make them likely to be structures local to the GC.
    \item  We find a range of relative orientations in the LFs as seen in Table \ref{tab:bfield_prop}. The relative orientation of the LFs and the FIREPLACE 214~$\mu$m magnetic field exhibits a bimodal distribution at 30\degree\ and 90\degree. These bimodal components reveal potential connections to the horizontal and vertical magnetic fields previously inferred in the region. This bimodal distribution disappears at 850~$\mu$m, where only a magnetic field component at 0\degree\ is observed as can be seen in the panels of the right column of Figure \ref{fig:hro_rht_jcmt}. 
    \item The comparison to the synthetic data sets generated from the MHD model described in Section \ref{sec:model} allows us to probe the properties of large filaments in magnetically dominated environments. We find that magnetically dominated filaments exhibit similar AM and HRO distributions to what is observed in LFs dominated by subsonic turbulence and shear. Subsonic turbulence and shear are possible mechanisms that could inhibit star formation. The result obtained from the synthetic observations therefore tentatively indicates that clouds and filaments in magnetically dominated environments could also be supported against collapse.
    \item We observe differences in the relative orientations of individual LFs between the 214 and 850~$\mu$m magnetic fields. These differences could be a result of complex magnetic field or dust geometry along the lines of sight towards these filaments. The 850~$\mu$m BISTRO observations also mostly cover LFs in the central region of the CMZ, and so the change in magnetic field orientation could indicate how the magnetic field structure changes throughout the GC.
    \item The different mechanisms dominating in the LFs indicate the changing role of turbulence where it is variously assisting in cloud collapse (supersonic turbulence and shock compression) and hindering gas compression (subsonic turbulence and shear). These results agree with previous studies that indicate that the physical conditions in the CMZ vary throughout the region \citep[e.g.][]{Pare2025}. These changing conditions help explain the range of SFRs seen in the CMZ, and in particular the generally low SFR seen in the region. 
    \item The LFs studied in this work are comparable to the ``bones'' studied in the Galactic Disk and we propose the label of ``ribs'' for these structures as the CMZ analogs of the Galactic Disk bones. The range of magnetic field orientations derived for the LFs studied in this work contrasts with the generally perpendicular field orientations derived for the Galactic Disk bones \citep{Pillai2015,Stephens2022,Coude2025}.
\end{itemize}

\facility{
    ALMA,
    SOFIA,
    JCMT
    }

\software{
    Astropy \citep{Astropy2022,Astropy2018,Astropy2013},
    Matplotlib \citep{Hunter2007}, 
    Numpy \citep{Harris2020},
    \textsc{polaris} \citep{Reissl2019}
    }

\textbf{Acknowledgments}

We would like to thank the anonymous referee for their valuable input and comments on this work. This work was written as part of a ``paper sprint'', in which a dedicated team of researchers engaged in an intense, two-week collaborative research and writing process. Please reach out to Dylan Par\'e for more details on the organization and outcome of this paper sprint. This paper makes use of the following ALMA data: ADS/JAO.ALMA\#2021.1.00172.S. ALMA is a partnership of ESO (representing its member states), NSF (USA) and NINS (Japan), together with NRC (Canada), NSTC and ASIAA (Taiwan), and KASI (Republic of Korea), in cooperation with the Republic of Chile. The Joint ALMA Observatory is operated by ESO, AUI/NRAO and NAOJ."
The author list is organized as follows: the PI, 5 co-leads (in alphabetical order), 7 other paper sprint participants (in alphabetical order), then all other contributing members of the ACES collaboration (in alphabetical order).

COOL Research DAO \citep{cool_whitepaper} is a Decentralized Autonomous Organization supporting research in astrophysics aimed at uncovering our cosmic origins.
D.P. acknowledges support from NASA ADAP award number 80NSSC25K7561

R.S.K. acknowledges financial support from the ERC via Synergy Grant ``ECOGAL'' (project ID 855130),  from the German Excellence Strategy via the Heidelberg Cluster ``STRUCTURES'' (EXC 2181 - 390900948), and from the German Ministry for Economic Affairs and Climate Action in project ``MAINN'' (funding ID 50OO2206). 

D.L. gratefully acknowledges funding from the National Science Foundation under Award Nos. 1816715, 2108938, and CAREER 2145689; as well as NASA FINESST Award No. 80NSSC24K1474.

C.F. acknowledges funding provided by the Australian Research Council (Discovery Project grants DP230102280 and DP250101526), and the Australia-Germany Joint Research Cooperation Scheme (UA-DAAD).

Z.F. and R.G.T. acknowledge support by the state of Baden-Württemberg through bwHPC
and the German Research Foundation (DFG) through grant INST 35/1597-1 FUGG.

A.G. acknowledges support from the NSF under grants AAG 2206511 and CAREER 2142300.

L.C. acknowledges support from the grant PID2022-136814NB-I00 by
the Spanish Ministry of Science, Innovation and Universities/State Agency of Research MICIU/AEI/10.13039/501100011033 and by ERDF, UE.

C.B.  gratefully  acknowledges  funding  from  National  Science  Foundation  under  Award  Nos. 2108938, 2206510, and CAREER 2145689, as well as from the National Aeronautics and Space Administration through the Astrophysics Data Analysis Program under Award ``3-D MC: Mapping Circumnuclear Molecular Clouds from X-ray to Radio,” Grant No. 80NSSC22K1125.

D.L.W gratefully acknowledges support from the UK ALMA Regional Centre (ARC) Node, which is supported by the Science and Technology Facilities Council (grant numbers ST/Y004108/1 and ST/T001488/1).

J.K. gratefully acknowledges support by the Royal Society under grant number RF\textbackslash ERE\textbackslash231132, as part of project URF\textbackslash R1\textbackslash211322.

Q.Z. gratefully acknowledges the support by the National Science Foundation under award No. AST-2206512 and the Smithsonian Institution FY2024 Scholarly Studies Program.


\bibliography{striations}{}
\bibliographystyle{aasjournal}

\end{document}